\documentclass[12pt,english]{article}
\usepackage[T1]{fontenc}
\usepackage[latin9]{inputenc}
\usepackage{verbatim}
\usepackage{amsmath}
\usepackage{amssymb}
\usepackage{graphicx}

\makeatletter
\numberwithin{equation}{section}

\makeatother

\usepackage{babel}
\begin{document}
\begin{titlepage}

\renewcommand{\thefootnote}{\fnsymbol{footnote}}

\begin{flushright}
\begin{tabular}{l}
UTHEP-644
\end{tabular}
\end{flushright}

\bigskip
\begin{center} {\Large \bf  Energy from the gauge invariant observables \\} \end{center}
\bigskip
\begin{center} 
{\large Takayuki Baba}\footnote{e-mail:      baba@het.ph.tsukuba.ac.jp} and {\large  Nobuyuki Ishibashi}\footnote{e-mail:         ishibash@het.ph.tsukuba.ac.jp} \end{center}
\begin{center} {\it Graduate School of Pure and Applied Sciences, University of Tsukuba,\\ Tsukuba, Ibaraki 305-8571, Japan} \end{center} 
\bigskip
\bigskip
\bigskip
\begin{abstract}  
For a classical solution $|\Psi \rangle $ in Witten's cubic string field theory, the gauge invariant observable $\langle I|\mathcal{V}|\Psi \rangle$ is conjectured to be equal to the difference of the one-point functions of the closed string state corresponding to $\mathcal{V}$, between the trivial vacuum and the one described by $|\Psi\rangle$. For a static solution $|\Psi\rangle$, if $\mathcal{V}$ is taken to be $c\bar{c}\partial X^0\bar{\partial}X^0$, the gauge invariant observable is expected to be proportional to the energy of $|\Psi\rangle$. We prove this relation assuming that $|\Psi \rangle $ satisfies equation of motion and some regularity conditions. We discuss how this relation can be applied to various solutions obtained recently.    \end{abstract}

\setcounter{footnote}{0} \renewcommand{\thefootnote}{\arabic{footnote}}

\end{titlepage}

\section{Introduction}

A great variety of analytic classical solutions have been found for
Witten's cubic string field theory \cite{Witten:1985cc}, since the
discovery of the analytic tachyon vacuum solution by Schnabl \cite{Schnabl:2005gv}
\footnote{For a review on these solutions, see \cite{Fuchs:2008cc}.%
}. In order to study the physical properties of these solutions, important
gauge invariant quantities to be calculated are the energy and the
gauge invariant observables $\left\langle I|\mathcal{V}\left(i\right)|\Psi\right\rangle $
discovered in \cite{Hashimoto:2001sm,Gaiotto:2001ji}. The gauge invariant
observable is conjectured to coincide with the difference of the one-point
functions of an on-shell closed string state between the trivial vacuum
and the one described by the solution $\left|\Psi\right\rangle $
\cite{Ellwood:2008jh,Kiermaier:2008qu}. 

What we would like to show in this paper is that energy can be expressed
by using a gauge invariant observable. Namely, for a static solution
$\left|\Psi\right\rangle $ of the equation of motion, the gauge invariant
observable with%
\footnote{Throughout this paper, we assume that the variable $X^{0}$ is described
by the free worldsheet theory with the Neumann boundary condition. %
} 
\begin{equation}
\mathcal{V}=\frac{2}{\pi i}c\bar{c}\partial X^{0}\bar{\partial}X^{0}\,,\label{eq:constantgraviton}
\end{equation}
is proportional to the energy:
\begin{equation}
E=\frac{1}{g^{2}}\left\langle I|\mathcal{V}\left(i\right)|\Psi\right\rangle \,.\label{eq:Egio}
\end{equation}
Here $g$ is the coupling constant of the string field theory. Naively
such a gauge invariant observable is proportional to the expectation
value of the energy momentum tensor $\left\langle T_{00}\right\rangle $
and thus the energy of the system. %
Usually, the energy is more difficult to calculate compared with the
gauge invariant observables. For most of the solutions obtained so
far, both the energy and the gauge invariant observable are calculated
and it turns out that the results are consistent with (\ref{eq:Egio}). 

In this paper, we will prove that (\ref{eq:Egio}) holds if $\left|\Psi\right\rangle $
satisfies the equation of motion and some regularity conditions. The
state-operator correspondence of the worldsheet theory implies that
the string field $\left|\Psi\right\rangle $ can be expressed as 
\[
\mathcal{O}_{\Psi}\left|0\right\rangle \,,
\]
where $\left|0\right\rangle $ is the SL(2,$\mathbb{R}$) invariant
vacuum and $\mathcal{O}_{\Psi}$ can be expressed in terms of local
operators on the upper half plane. We will first discuss the case
in which $\mathcal{O}_{\Psi}$ consists of local operators located
away from the curve $\left|\xi\right|=1$, where $\xi$ is the complex
coordinate on the upper half plane. As we will see, the proof of (\ref{eq:Egio})
is relatively easy in such a case. However, most of the solutions
obtained since \cite{Schnabl:2005gv} do not satisfy this condition
because they involve non-local operators such as $K,B$. Fortunately
our method of proof can be refined to be applicable to such cases.
We discuss applications of our results to the solutions obtained recently. 

This paper is organized as follows. In section \ref{sec:derivation},
we give a proof of the relation (\ref{eq:Egio}), assuming $\left|\Psi\right\rangle $
can be expressed using local operators. In section \ref{sec:KBc},
we take Okawa type solutions \cite{Okawa:2006vm,Erler:2006hw,Erler:2006ww}
as an example and explain how we should generalize our method of proof
to deal with solutions involving non-local operators $K,B$. In section
\ref{sec:Other-solutions} we apply our results to other solutions
discovered recently. Section \ref{sec:Conclusion-and-discussion}
is devoted to discussions.

\section{A proof of (\ref{eq:Egio}) for local $\mathcal{O}_{\Psi}$\label{sec:derivation}}

In this section, we consider the case in which $\mathcal{O}_{\Psi}$
is made from local operators located away from $\left|\xi\right|=1$.
We also assume that $\mathcal{O}_{\Psi}$ does not involve $X^{0}$
variable.

\subsection{Open string field theory in a weak gravitational background}

In order to derive (\ref{eq:Egio}), we start from considering the
following modification of the string field action,
\begin{equation}
S_{h}=-\frac{1}{g^{2}}\left[\frac{1}{2}\left\langle \Psi|Q|\Psi\right\rangle +\frac{1}{3}\left\langle \Psi|\Psi*\Psi\right\rangle +h\left\langle I|\mathcal{V}\left(i\right)|\Psi\right\rangle \right]\,,\label{eq:Sh}
\end{equation}
with $h\ll1$. It has been shown in \cite{Zwiebach:1992bw} that such
a string field action describes string theory in a closed string background,
for general on-shell $\mathcal{V}$. %
The vertex operator $\mathcal{V}$ in (\ref{eq:constantgraviton})
is a linear combination of those for the constant graviton and dilaton.
Therefore the action (\ref{eq:Sh}) should be the open string field
theory in a constant metric and dilaton background. 

\begin{figure}[b]
\begin{centering}
\includegraphics[scale=0.5]{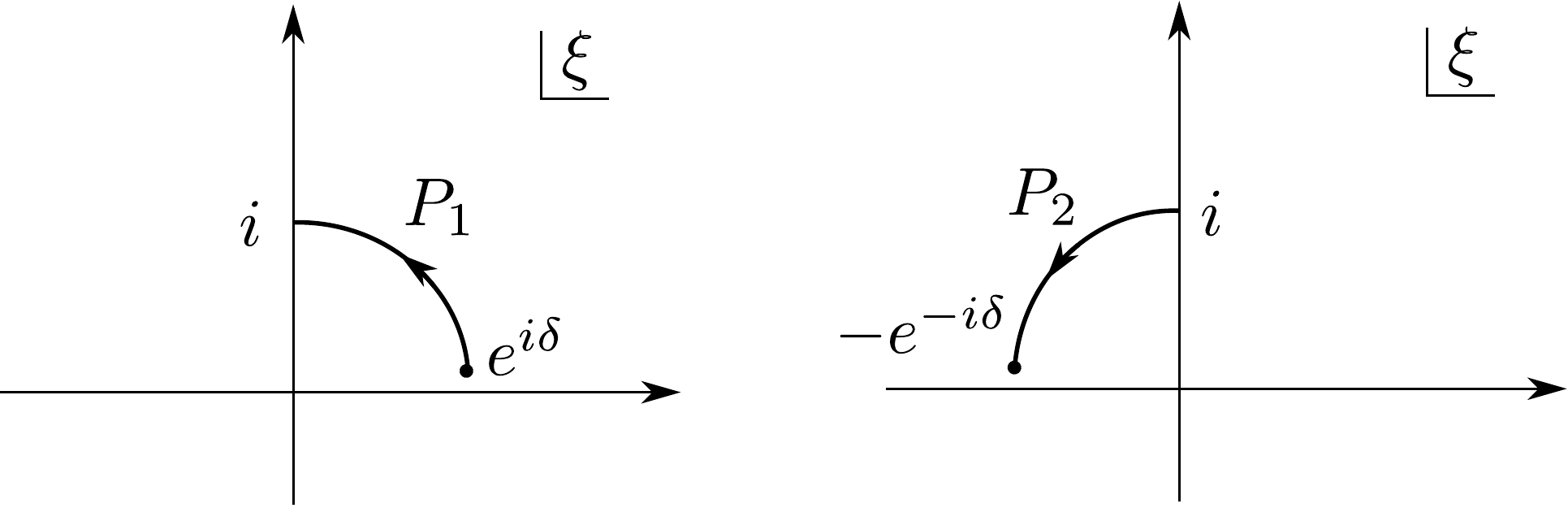}
\par\end{centering}

\caption{\label{fig:Contours}Contours $P_{1},P_{2}$}
\end{figure}

By a general coordinate transformation, the constant metric can be
turned into the original $\eta_{\mu\nu}$. Therefore we expect that
we can somehow transform the string field action (\ref{eq:Sh}) into
the original string field action with some rescaling of the coupling
constant $g$. In order to do so, we notice that as an operator acting
on $\mathcal{O}_{\Psi}\left|0\right\rangle $, $\mathcal{V}$ can
be expressed in a BRST exact form
\begin{equation}
\mathcal{V}\left(i\right)=\left\{ Q,\chi\right\} \,,\label{eq:Qchi}
\end{equation}
where
\begin{eqnarray}
\chi & \equiv & \lim_{\delta\to0}\left[\int_{P_{1}}\frac{d\xi}{2\pi i}j\left(\xi,\bar{\xi}\right)-\int_{\bar{P}_{1}}\frac{d\bar{\xi}}{2\pi i}\bar{j}\left(\xi,\bar{\xi}\right)+\frac{c\left(1\right)}{2\pi\delta}\right]\,,\label{eq:chi}\\
 &  & j\left(\xi,\bar{\xi}\right)\equiv4\partial X^{0}\left(\xi\right)\bar{c}\bar{\partial}X^{0}\left(\bar{\xi}\right)\,,\nonumber \\
 &  & \bar{j}\left(\xi,\bar{\xi}\right)\equiv4\bar{\partial}X^{0}\left(\bar{\xi}\right)c\partial X^{0}\left(\xi\right)\,.\nonumber 
\end{eqnarray}
Here $P_{1}$ is the path depicted in Fig \ref{fig:Contours} and
along the arcs of the circle $\left|\xi\right|=1$. Because of our
assumption, the presence of $\mathcal{O}_{\Psi}$ does not affect
the operators defined on such contours. %
{} Since $j,\bar{j}$ diverge in the limit $\mathrm{Im}\xi\to0$, we
have introduced $\delta>0$ to regularize the divergence. One can
check that the limit on the right hand side of (\ref{eq:chi}) is
not singular. 

(\ref{eq:Qchi}) implies that in terms of the string field $\left|\Psi^{\prime}\right\rangle $
defined as
\begin{equation}
\left|\Psi^{\prime}\right\rangle \equiv\left|\Psi\right\rangle +h\chi\left|I\right\rangle \,,\label{eq:Psiprime}
\end{equation}
 the string field action $S_{h}$ is expressed as 
\begin{equation}
S_{h}=-\frac{1}{g^{2}}\left[\frac{1}{2}\left\langle \Psi^{\prime}|Q^{\prime}|\Psi^{\prime}\right\rangle +\frac{1}{3}\left\langle \Psi^{\prime}|\Psi^{\prime}*\Psi^{\prime}\right\rangle \right]+\mathcal{O}\left(h^{2}\right)\,,\label{eq:ShQprime}
\end{equation}
with
\begin{eqnarray*}
Q^{\prime} & \equiv & Q-h\left(\chi-\chi^{\dagger}\right)\,.
\end{eqnarray*}
$\chi^{\dagger}$ denotes the BPZ conjugate of $\chi$ and
\begin{eqnarray*}
\chi-\chi^{\dagger} & = & \lim_{\delta\to0}\left[\int_{P_{1}+P_{2}}\frac{d\xi}{2\pi i}j\left(\xi,\bar{\xi}\right)-\int_{\bar{P}_{1}+\bar{P}_{2}}\frac{d\bar{\xi}}{2\pi i}\bar{j}\left(\xi,\bar{\xi}\right)\right.\\
 &  & \hphantom{Q-h\lim_{\delta\to0}\quad\int_{C_{1}+C_{2}}\frac{d\xi}{2\pi i}}\left.+\frac{c\left(1\right)}{2\pi\delta}-\frac{c\left(-1\right)}{2\pi\delta}\right]\,,
\end{eqnarray*}
where $P_{2}$ is the contour depicted in Fig. \ref{fig:Contours}.
We give the details of the definition of $\chi$ and the derivation
of (\ref{eq:Qchi})(\ref{eq:ShQprime}) in appendix \ref{sec:Derivation-of-}. 

\begin{figure}
\begin{centering}
\includegraphics[scale=0.6]{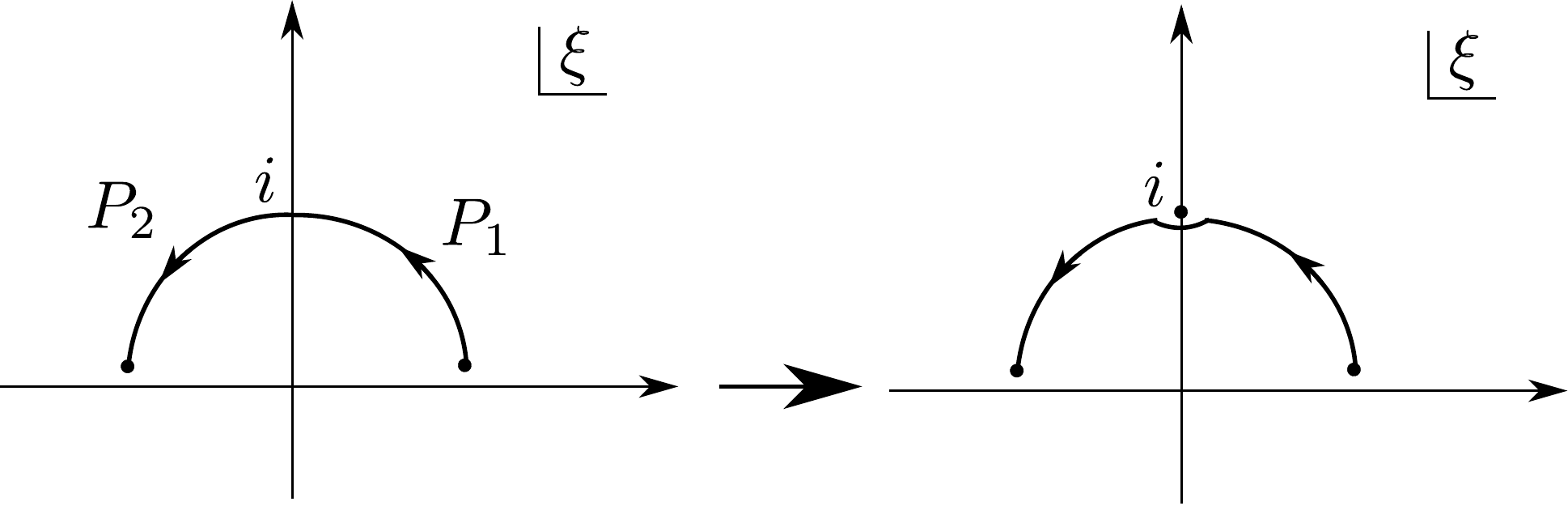}
\par\end{centering}

\caption{the contour to define $\mathcal{G}$\label{fig:the-contour-to}}
\end{figure}

Therefore the string field theory in the weak %
{} background is given by the cubic action with the modified BRST operator
$Q^{\prime}$. This string field theory is similar to the one considered
in \cite{Myers:1988zw} as the open string field theory in the soft
dilaton background. They have shown that the effect of such a background
corresponds to a rescaling of the string coupling constant $g$. It
is straightforward to generalize the techniques of \cite{Myers:1988zw}
to our case. Let us define
\begin{eqnarray}
\mathcal{G} & \equiv & \lim_{\delta\to0}\left[\int_{P_{1}+P_{2}}\frac{d\xi}{2\pi i}g_{\xi}\left(\xi,\bar{\xi}\right)-\int_{\bar{P}_{1}+\bar{P}_{2}}\frac{d\bar{\xi}}{2\pi i}g_{\bar{\xi}}\left(\xi,\bar{\xi}\right)\right]\,,\label{eq:mathcalG}\\
 &  & g_{\xi}\left(\xi,\bar{\xi}\right)\equiv2\left(X^{0}\left(\xi,\bar{\xi}\right)-X^{0}\left(i,-i\right)\right)\partial X^{0}\left(\xi\right)\,,\nonumber \\
 &  & g_{\bar{\xi}}\left(\xi,\bar{\xi}\right)\equiv2\left(X^{0}\left(\xi,\bar{\xi}\right)-X^{0}\left(i,-i\right)\right)\bar{\partial}X^{0}\left(\bar{\xi}\right)\,.\nonumber 
\end{eqnarray}
Because of the presence of $X^{0}\left(i,-i\right)$, $g_{\xi},g_{\bar{\xi}}$
are well-defined operators on the worldsheet. Since $g_{\xi},g_{\bar{\xi}}$
are singular at $\xi=i$, on the right hand side of (\ref{eq:mathcalG})
the integration contour is modified infinitesimally as in Figure \ref{fig:the-contour-to}.
$g_{\xi},g_{\bar{\xi}}$ are defined with the usual normal ordering
prescription (\ref{eq:normalorder}) and under a conformal transformation
$\xi\to\xi^{\prime}\left(\xi\right)$, $g_{\xi}$ transforms as 
\begin{equation}
g_{\xi^{\prime}}\left(\xi^{\prime},\bar{\xi}^{\prime}\right)=\frac{\partial\xi}{\partial\xi^{\prime}}g_{\xi}\left(\xi,\bar{\xi}\right)+\frac{1}{2}\partial_{\xi^{\prime}}\ln\frac{\partial\xi}{\partial\xi^{\prime}}\,.\label{eq:gxiprime}
\end{equation}
It is straightforward to check that the limit on the right hand side
of (\ref{eq:mathcalG}) is not singular.

Using (\ref{eq:gxiprime}) and the fact that $g_{\xi},g_{\bar{\xi}}$
are singular at $\xi=i$, one can deduce the following identities:

\begin{eqnarray}
 &  & \left\langle \mathcal{G}\Psi_{1}|\Psi_{2}\right\rangle +\left\langle \Psi_{1}|\mathcal{G}\Psi_{2}\right\rangle =\left\langle \Psi_{1}|\Psi_{2}\right\rangle \,,\label{eq:mathcalGGdagger}\\
 &  & \left\langle \mathcal{G}\Psi_{1}|\Psi_{2}\ast\Psi_{3}\right\rangle +\left\langle \Psi_{1}|\mathcal{G}\Psi_{2}\ast\Psi_{3}\right\rangle +\left\langle \Psi_{1}|\Psi_{2}\ast\mathcal{G}\Psi_{3}\right\rangle =\left\langle \Psi_{1}|\Psi_{2}\ast\Psi_{3}\right\rangle \,.\label{eq:mathcalGABC}
\end{eqnarray}
As is explained in appendix \ref{sec:Derivation-of-}, it is also
straightforward to get 
\begin{equation}
\left[Q,\mathcal{G}\right]=\chi-\chi^{\dagger}\,.\label{eq:QmathcalG}
\end{equation}
Then, in terms of 
\[
\left|\Psi^{\prime\prime}\right\rangle \equiv\left(1-h\mathcal{G}\right)\left|\Psi^{\prime}\right\rangle \,,
\]
$S_{h}$ can be expressed as
\begin{equation}
S_{h}=-\frac{1+h}{g^{2}}\left[\frac{1}{2}\left\langle \Psi^{\prime\prime}|Q|\Psi^{\prime\prime}\right\rangle +\frac{1}{3}\left\langle \Psi^{\prime\prime}|\Psi^{\prime\prime}*\Psi^{\prime\prime}\right\rangle \right]+\mathcal{O}\left(h^{2}\right)\,.\label{eq:Sh2}
\end{equation}
Thus $S_{h}$ is proportional to the original string field theory
action for the string field $\left|\Psi^{\prime\prime}\right\rangle $.
By a field redefinition, the effect of the weak %
{} background is turned into a rescaling of the coupling constant $g$,
due to the constant dilaton background. $\mathcal{G}$ can be regarded
as the generator of general coordinate transformation. 

\subsection{Derivation of (\ref{eq:Egio})}

We can derive (\ref{eq:Egio}) from the two expressions (\ref{eq:Sh})(\ref{eq:Sh2})
of $S_{h}$. Suppose that $\left|\Psi\right\rangle $ is a static
solution of the equation of motion
\[
Q\left|\Psi\right\rangle +\left|\Psi\ast\Psi\right\rangle =0\,,
\]
and evaluate $S_{h}$ using eqs.(\ref{eq:Sh})(\ref{eq:Sh2}). The
right hand side of (\ref{eq:Sh}) can be expressed as 
\[
S_{h}=-E-\frac{h}{g^{2}}\left\langle I|\mathcal{V}\left(i\right)|\Psi\right\rangle \,,
\]
where $E=\frac{1}{g^{2}}\left[\frac{1}{2}\left\langle \Psi|Q|\Psi\right\rangle +\frac{1}{3}\left\langle \Psi|\Psi*\Psi\right\rangle \right]$
is the energy of the solution $\left|\Psi\right\rangle $. On the
other hand, since $\left|\Psi^{\prime\prime}\right\rangle $ can be
expressed as
\[
\left|\Psi^{\prime\prime}\right\rangle =\left|\Psi\right\rangle +\left|\delta^{\prime\prime}\Psi\right\rangle \,,
\]
with $\left|\delta^{\prime\prime}\Psi\right\rangle \sim\mathcal{O}\left(h\right)$,
the right hand side of (\ref{eq:Sh2}) becomes 
\begin{eqnarray*}
 &  & -\frac{1+h}{g^{2}}\left[\frac{1}{2}\left\langle \Psi|Q|\Psi\right\rangle +\frac{1}{3}\left\langle \Psi|\Psi*\Psi\right\rangle \right.\\
 &  & \hphantom{-\frac{1+h}{g^{2}}\quad}\left.+\left\langle \delta^{\prime\prime}\Psi\right|\left(Q\left|\Psi\right\rangle +\left|\Psi\ast\Psi\right\rangle \right)\right]+\mathcal{O}\left(h^{2}\right)\,.
\end{eqnarray*}
Using the fact that $\left|\Psi\right\rangle $ is a solution of the
equation of motion, one can see that (\ref{eq:Sh2}) can be rewritten
as 
\[
S_{h}=-\left(1+h\right)E+\mathcal{O}\left(h^{2}\right)\,.
\]
Comparing these, we obtain
\begin{equation}
E=\frac{1}{g^{2}}\left\langle I|\mathcal{V}\left(i\right)|\Psi\right\rangle \,.\label{eq:Egio3}
\end{equation}

There is a more direct way to derive (\ref{eq:Egio3}), which is essentially
equivalent to the one above and will be used in the subsequent sections.
From (\ref{eq:mathcalGGdagger})(\ref{eq:mathcalGABC}), one can deduce
\begin{eqnarray}
\frac{1}{3}\left\langle \Psi|\Psi\ast\Psi\right\rangle  & = & \left\langle \mathcal{G}\Psi|\Psi\ast\Psi\right\rangle \,,\nonumber \\
\frac{1}{2}\left\langle \Psi\right|Q\left|\Psi\right\rangle  & = & \left\langle \mathcal{G}\Psi\right|Q\left|\Psi\right\rangle -\frac{1}{2}\left\langle \Psi\right|\left[Q,\mathcal{G}\right]\left|\Psi\right\rangle \,,\label{eq:mathcalGPsiQPsi}
\end{eqnarray}
and from (\ref{eq:QmathcalG}) we get 
\begin{equation}
\left[Q,\mathcal{G}\right]\left|\Psi\right\rangle =\left(\chi-\chi^{\dagger}\right)\left|\Psi\right\rangle \,.\label{eq:QmathcalGPsi}
\end{equation}
Using these and the equation of motion, we obtain
\begin{eqnarray}
E & = & \frac{1}{g^{2}}\left[\frac{1}{2}\left\langle \Psi|Q|\Psi\right\rangle +\frac{1}{3}\left\langle \Psi|\Psi*\Psi\right\rangle \right]\nonumber \\
 & = & \frac{1}{g^{2}}\left[\left\langle \mathcal{G}\Psi\right|\left\{ Q\left|\Psi\right\rangle +\left|\Psi\ast\Psi\right\rangle \right\} -\frac{1}{2}\left\langle \Psi\right|\left[Q,\mathcal{G}\right]\left|\Psi\right\rangle \right]\nonumber \\
 & = & -\frac{1}{2g^{2}}\left\langle \Psi\right|\left(\chi-\chi^{\dagger}\right)\left|\Psi\right\rangle \nonumber \\
 & = & -\frac{1}{g^{2}}\left\langle I\right|\chi\left|\Psi\ast\Psi\right\rangle \nonumber \\
 & = & \frac{1}{g^{2}}\left\langle I\right|\chi Q\left|\Psi\right\rangle \nonumber \\
 & = & \frac{1}{g^{2}}\left\langle I|\mathcal{V}\left(i\right)|\Psi\right\rangle \,.\label{eq:Egio2}
\end{eqnarray}
\medskip{}

Before closing this section, a few comments are in order:%

\begin{itemize}
\item The vertex operator $\mathcal{V}$ is expressed in a BRST exact form
(\ref{eq:Qchi}), with $\chi$ being a completely legal operator.
This fact may appear odd because it implies that all the amplitudes
involving $\mathcal{V}$ vanish%
\footnote{This question was raised by M. Schnabl. %
}. Actually (\ref{eq:Qchi}) holds on the assumption that there exists
no operators around $\xi=1$. In the derivation of (\ref{eq:Qchi})
in appendix \ref{sec:Derivation-of-}, we use (\ref{eq:Veidelta})
which is valid only when such a condition is satisfied, which is the
case in our setup. However, in calculating amplitudes, this is not
guaranteed because of the existence of other vertex operators and
(\ref{eq:Qchi}) cannot be used in such a situation. 
\item It is also possible to use %
\[
\mathcal{V}=c\bar{c}\partial X^{\mu}\bar{\partial}X^{\nu}h_{\mu\nu}\,,
\]
with $h_{\mu}^{\mu}=-1$ and derive (\ref{eq:Egio}), provided the
variables $X^{\mu}$ are described by the free worldsheet theory with
the Neumann boundary condition.
\item Suppose that $\left|\Psi\right\rangle $ does not satisfy the equation
of motion:
\begin{equation}
Q\left|\Psi\right\rangle +\left|\Psi\ast\Psi\right\rangle \equiv\left|\Gamma\right\rangle \ne0\,.\label{eq:Gamma}
\end{equation}
It is easy to see that the relation (\ref{eq:Egio2}) is modified
as 
\begin{equation}
E=\frac{1}{g^{2}}\left\langle I|\mathcal{V}\left(i\right)|\Psi\right\rangle -\frac{1}{g^{2}}\left\langle I\right|\chi\left|\Gamma\right\rangle +\frac{1}{g^{2}}\left\langle \mathcal{G}\Psi|\Gamma\right\rangle \,.\label{eq:anomaly}
\end{equation}

\end{itemize}

\section{Derivation of (\ref{eq:Egio}) for Okawa type solutions\label{sec:KBc}}

{} Most of the nontrivial solutions obtained so far are described by
using operators $K,B$. These operators are given as integrations
of $T,b$ along the contours which intersects $P_{1},\bar{P}_{1},P_{2},\bar{P}_{2}$
and do not commute with $g_{\xi},g_{\bar{\xi}}$ to be used to define
$\mathcal{G}$. In order to prove (\ref{eq:Egio}) for such $\left|\Psi\right\rangle ,$
we need to define the quantities which appear in the previous section
in the presence of such operators. Moreover it is not so straightforward
to prove (\ref{eq:QmathcalGPsi}) in such a setup. 

In this section, as a prototype of such solutions, we consider the
Okawa type solutions \cite{Okawa:2006vm,Erler:2006hw,Erler:2006ww}
\begin{equation}
\Psi=F\left(K\right)c\frac{KB}{1-F\left(K\right)^{2}}cF\left(K\right)\,.\label{eq:Okawa}
\end{equation}
Here $\Psi$ is expressed in terms of string fields $K,B,c$ and the
product of them is the star product%
\footnote{See \cite{Erler:2009uj,Schnabl:2010tb} for details.%
}. $\Psi$ gives a solution of the equation of motion if $F\left(K\right),\frac{K}{1-F^{2}}$
are sufficiently regular functions of $K$. We will show that it is
possible to define $\mathcal{G}$ which acts on such solutions and
prove (\ref{eq:mathcalGPsiQPsi})(\ref{eq:QmathcalGPsi}) and derive
(\ref{eq:Egio2}). 

It is assumed that $F\left(K\right),\frac{K}{1-F^{2}}$ are given
in a Laplace transformed form 
\begin{eqnarray*}
F\left(K\right) & = & \int_{0}^{\infty}dLe^{-LK}f\left(L\right)\,,\\
\frac{K}{1-F^{2}} & = & \int_{0}^{\infty}dLe^{-LK}\tilde{f}\left(L\right)\,.
\end{eqnarray*}
Substituting these into (\ref{eq:Okawa}), we obtain an expression
of $\Psi$ 
\begin{equation}
\Psi=\int_{0}^{\infty}dLe^{-LK}\psi\left(L\right)\,,\label{eq:wedge}
\end{equation}
where
\begin{eqnarray}
\psi\left(L\right) & = & \int dL_{1}dL_{2}dL_{3}\delta\left(L-L_{1}-L_{2}-L_{3}\right)\nonumber \\
 &  & \quad\times c\left(L_{2}+L_{3}\right)Bc\left(L_{3}\right)f\left(L_{1}\right)\tilde{f}\left(L_{2}\right)f\left(L_{3}\right)\,,\label{eq:Psitilde}
\end{eqnarray}
and
\begin{equation}
c\left(z\right)=e^{zK}ce^{-zK}\,.\label{eq:c(t)}
\end{equation}
$\Psi$ can be considered as the Laplace transform of $\psi$. We
express (\ref{eq:wedge}) as 
\[
\Psi=\mathcal{L}\left\{ \psi\right\} \,,
\]
where $\mathcal{L}$ denotes the operation of the Laplace transform.
Then $\psi\left(L\right)$ is expressed as 
\[
\psi\left(L\right)=\mathcal{L}^{-1}\left\{ \Psi\right\} \left(L\right)\,.
\]

\subsection{Definition of $\mathcal{G}$ }

\begin{figure}
\begin{centering}
\includegraphics[scale=0.8]{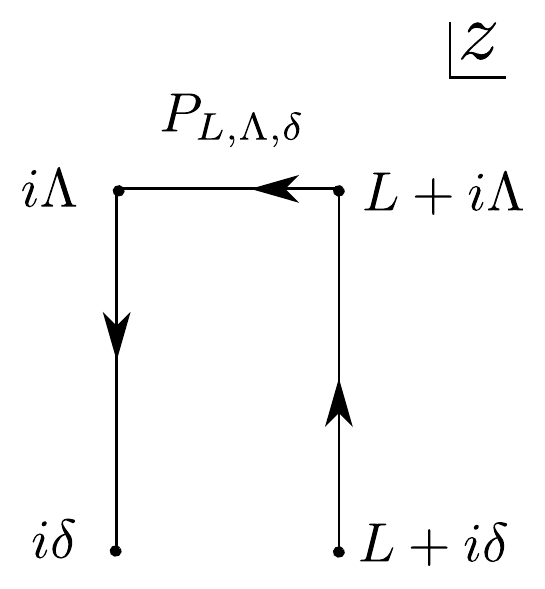}
\par\end{centering}

\caption{$P_{L,\Lambda,\delta}$\label{fig:PLLambdadelta}}
\end{figure}

$\Psi$ is represented as a sum of wedge states with insertions $e^{-LK}\psi\left(L\right)$
as (\ref{eq:wedge}). In order to define $\mathcal{G}$ which acts
on such $\Psi$, the contour to be used should depend on the length
$L$ of the wedge state. So we introduce 
\begin{eqnarray*}
\mathcal{G}(L,\Lambda,\delta) & \equiv & \lim_{z_{0}\to i\infty}\left[\int_{P_{L,\Lambda,\delta}}\frac{dz}{2\pi i}g_{z}(z,\bar{z})-\int_{\bar{P}_{L,\Lambda,\delta}}\frac{d\bar{z}}{2\pi i}g_{\bar{z}}(z,\bar{z})\right]\,,\\
 &  & g_{z}(z,\bar{z})=2\left(X^{0}\left(z,\bar{z}\right)-X^{0}(z_{0},\bar{z}_{0})\right)\partial X^{0}(z)\,,\\
 &  & g_{\bar{z}}(z,\bar{z})=2\left(X^{0}\left(z,\bar{z}\right)-X^{0}(z_{0},\bar{z}_{0})\right)\bar{\partial}X^{0}(z)\,,
\end{eqnarray*}
and define $\mathcal{G}\Psi$ so that for any test state $\left|\phi\right\rangle =\phi\left(0\right)\left|0\right\rangle $,
$\left\langle \phi|\mathcal{G}\Psi\right\rangle $ is given as 
\begin{eqnarray}
\left\langle \phi|\mathcal{G}\Psi\right\rangle  & = & \lim_{\left(\Lambda,\delta\right)\to\left(\infty,0\right)}\int_{0}^{\infty}dL\left\langle e^{\left(L+\frac{1}{2}\right)K}f\circ\phi\left(0\right)e^{-\left(L+\frac{1}{2}\right)K}\mathcal{G}(L,\Lambda,\delta)\psi\left(L\right)\right\rangle _{C_{L+1}}\,.\label{eq:mathcalGPsi}
\end{eqnarray}
 %
Here $f\left(\xi\right)\equiv\frac{\pi}{2}\arctan\xi$, $\left\langle \cdots\right\rangle _{C_{L+1}}$
denotes the correlation function on the infinite cylinder $C_{L+1}$
with circumference $L+1$ and $\phi\left(0\right)$ in the correlation
function denotes the operator on $C_{L+1}$ corresponding to $|\phi\rangle$,
by abuse of notation. $z$ which appears in the definition of $\mathcal{G}(L,\Lambda,\delta)$
is the complex coordinate on $C_{L+1}$ such that $e^{-LK}\psi\left(L\right)$
corresponds to the region $0\leq\mathrm{Re}z\leq L$. The contour
$P_{L,\Lambda,\delta}$ is the one depicted in Figure (\ref{fig:PLLambdadelta}),
which consists of straight lines. 

With $\mathcal{G}$ thus defined, we will prove the identity (\ref{eq:mathcalGABC})
assuming $\Psi_{i}\,\left(i=1,2,3\right)$ does not involve the $X^{0}$
variable%
. $\left\langle \mathcal{G}\Psi_{1}|\Psi_{2}\ast\Psi_{3}\right\rangle $
is given as 
\begin{eqnarray}
 &  & \lim_{\left(\Lambda,\delta\right)\to\left(\infty,0\right)}\int_{0}^{\infty}dL_{1}\int_{0}^{\infty}dL_{2}\int_{0}^{\infty}dL_{3}\nonumber \\
 &  & \hphantom{\lim_{\left(\Lambda,\delta\right)\to\left(\infty,0\right)}\int dL_{1}}\times\left\langle e^{(L_{2}+L_{3})K}\mathcal{G}(L_{1},\Lambda,\delta)\psi_{1}(L_{1})e^{-L_{2}K}\psi_{2}(L_{2})e^{-L_{3}K}\psi_{3}(L_{3})\right\rangle _{C_{L_{1}+L_{2}+L_{3}}}\,,\nonumber \\
 &  & \,\label{eq:mathcalA}
\end{eqnarray}
in terms of the correlation function on the infinite cylinder $C_{L_{1}+L_{2}+L_{3}}$
with circumference $L_{1}+L_{2}+L_{3}$. Since $\psi_{i}$ does not
involve the $X^{0}$ variable, the correlation function on the right
hand side of (\ref{eq:mathcalA}) is factorized as 
\begin{eqnarray}
 &  & \left\langle e^{(L_{2}+L_{3})K}\mathcal{G}(L_{1},\Lambda,a)\psi_{1}(L_{1})e^{-L_{2}K}\psi_{2}(L_{2})e^{-L_{3}K}\psi_{3}(L_{3})\right\rangle _{C_{L_{1}+L_{2}+L_{3}}}\nonumber \\
 &  & \quad=\left\langle \mathcal{G}(L_{1},\Lambda,a)\right\rangle {}_{C_{L_{1}+L_{2}+L_{3}}}^{X^{0}}\nonumber \\
 &  & \hphantom{\quad=\quad}\times\left\langle e^{(L_{2}+L_{3})K}\psi_{1}(L_{1})e^{-L_{2}K}\psi_{2}(L_{2})e^{-L_{3}K}\psi_{3}(L_{3})\right\rangle {}_{C_{L_{1}+L_{2}+L_{3}}}\,,\label{eq:DPsiPsi}
\end{eqnarray}
where $\left\langle \cdots\right\rangle _{C_{L_{1}+L_{2}+L_{3}}}^{X^{0}}$
denotes the correlation function with respect to $X^{0}$ variable
on $C_{L_{1}+L_{2}+L_{3}}$. The expectation value $\langle\mathcal{G}(L_{1},\Lambda,\delta)\rangle_{C_{L_{1}+L_{2}+L_{3}}}^{X^{0}}$
can be calculated using (\ref{eq:X-X0CL}). In the limit $\Lambda\to\infty,\delta\to0$,
we obtain
\begin{equation}
\lim_{\left(\Lambda,\delta\right)\to\left(\infty,0\right)}\left\langle \mathcal{G}(L_{1},\Lambda,\delta)\right\rangle {}_{C_{L_{1}+L_{2}+L_{3}}}^{X^{0}}=\frac{L_{1}}{L_{1}+L_{2}+L_{3}}\,.\label{eq:expmathcalG}
\end{equation}
Therefore we get 
\begin{eqnarray*}
\left\langle \mathcal{G}\Psi_{1}|\Psi_{2}\ast\Psi_{3}\right\rangle  & = & \int dL_{1}dL_{2}dL_{3}\frac{L_{1}}{L_{1}+L_{2}+L_{3}}\\
 &  & \hphantom{\int dL_{1}dL_{2}}\times\left\langle e^{(L_{2}+L_{3})K}\psi_{1}(L_{1})e^{-L_{2}K}\psi_{2}(L_{2})e^{-L_{3}K}\psi_{3}(L_{3})\right\rangle {}_{C_{L_{1}+L_{2}+L_{3}}}\,.
\end{eqnarray*}
In the same way, we obtain 
\begin{eqnarray*}
\left\langle \Psi_{1}|\mathcal{G}\Psi_{2}\ast\Psi_{3}\right\rangle  & = & \int dL_{1}dL_{2}dL_{3}\frac{L_{2}}{L_{1}+L_{2}+L_{3}}\\
 &  & \hphantom{\int dL_{1}dL_{2}}\times\left\langle e^{(L_{2}+L_{3})K}\psi_{1}(L_{1})e^{-L_{2}K}\psi_{2}(L_{2})e^{-L_{3}K}\psi_{3}(L_{3})\right\rangle {}_{C_{L_{1}+L_{2}+L_{3}}}\,,\\
\left\langle \Psi_{1}|\Psi_{2}\ast\mathcal{G}\Psi_{3}\right\rangle  & = & \int dL_{1}dL_{2}dL_{3}\frac{L_{3}}{L_{1}+L_{2}+L_{3}}\\
 &  & \hphantom{\int dL_{1}dL_{2}}\times\left\langle e^{(L_{2}+L_{3})K}\psi_{1}(L_{1})e^{-L_{2}K}\psi_{2}(L_{2})e^{-L_{3}K}\psi_{3}(L_{3})\right\rangle {}_{C_{L_{1}+L_{2}+L_{3}}}\,,
\end{eqnarray*}
and from these (\ref{eq:mathcalGABC}) is obvious. (\ref{eq:mathcalGGdagger})
can also be proved in a similar way.

\subsection{(\ref{eq:QmathcalGPsi}) for Okawa type solutions }

Since (\ref{eq:mathcalGGdagger})(\ref{eq:mathcalGABC}) are satisfied,
(\ref{eq:mathcalGPsiQPsi}) can be deduced immediately. However, with
the definition of $\mathcal{G}$ given in (\ref{eq:mathcalGPsi}),
proving (\ref{eq:QmathcalGPsi}) is not so straightforward. We elaborate
on this here. 

From the definition (\ref{eq:mathcalGPsi}), we obtain
\begin{eqnarray*}
\left\langle \phi\right|\left[Q,\mathcal{G}\right]\left|\Psi\right\rangle  & = & \lim_{\left(\Lambda,\delta\right)\to\left(\infty,0\right)}\left[\int_{0}^{\infty}dL\left\langle e^{\left(L+\frac{1}{2}\right)K}f\circ\phi\left(0\right)e^{-\left(L+\frac{1}{2}\right)K}Q\mathcal{G}(L,\Lambda,\delta)\psi\left(L\right)\right\rangle _{C_{L+1}}\right.\\
 &  & \hphantom{\lim_{\left(\Lambda,\delta\right)\to\left(\infty,0\right)}\quad}\left.-\int_{0}^{\infty}dL\left\langle e^{\left(L+\frac{1}{2}\right)K}f\circ\phi\left(0\right)e^{-\left(L+\frac{1}{2}\right)K}\mathcal{G}(L,\Lambda,\delta)\mathcal{L}^{-1}\left\{ Q\Psi\right\} \left(L\right)\right\rangle _{C_{L+1}}\right]\\
 & = & \mathcal{A}_{1}+\mathcal{A}_{2}\,,
\end{eqnarray*}
where
\begin{eqnarray}
\mathcal{A}_{1} & \equiv & \lim_{\left(\Lambda,\delta\right)\to\left(\infty,0\right)}\int_{0}^{\infty}dL\left\langle e^{\left(L+\frac{1}{2}\right)K}f\circ\phi\left(0\right)e^{-\left(L+\frac{1}{2}\right)K}\left[Q,\mathcal{G}(L,\Lambda,\delta)\right]\psi\left(L\right)\right\rangle _{C_{L+1}}\,,\label{eq:mathcalA1}\\
\mathcal{A}_{2} & \equiv & \lim_{\left(\Lambda,\delta\right)\to\left(\infty,0\right)}\int_{0}^{\infty}dL\left\langle e^{\left(L+\frac{1}{2}\right)K}f\circ\phi\left(0\right)e^{-\left(L+\frac{1}{2}\right)K}\right.\nonumber \\
 &  & \hphantom{\lim_{\left(\Lambda,\delta\right)\to\left(\infty,0\right)}\int_{0}^{\infty}dL\quad e^{\left(L+\frac{1}{2}\right)K}}\times\left.\vphantom{e^{\left(L+\frac{1}{2}\right)K}f\circ\phi\left(0\right)e^{-\left(L+\frac{1}{2}\right)K}}\mathcal{G}(L,\Lambda,\delta)\left[Q\psi\left(L\right)-\mathcal{L}^{-1}\left\{ Q\Psi\right\} \left(L\right)\right]\right\rangle _{C_{L+1}}\,.\nonumber \\
 &  & \ \label{eq:mathcalA2}
\end{eqnarray}

Substituting (\ref{eq:Psitilde}) into (\ref{eq:mathcalA1}), we obtain
\begin{eqnarray}
\mathcal{A}_{1} & = & \int dL\int dL_{1}dL_{2}dL_{3}\delta\left(L-L_{1}-L_{2}-L_{3}\right)\nonumber \\
 &  & \hphantom{-\int dL_{1}dL_{2}}\times f\left(L_{1}\right)\tilde{f}\left(L_{2}\right)f\left(L_{3}\right)\nonumber \\
 &  & \hphantom{-\int dL_{1}dL_{2}}\times\left\langle e^{\left(L+\frac{1}{2}\right)K}f\circ\phi\left(0\right)e^{-\left(L+\frac{1}{2}\right)K}\left[Q,\mathcal{G}(L,\Lambda,\delta)\right]c\left(L_{2}+L_{3}\right)Bc\left(L_{3}\right)\right\rangle _{C_{L+1}}\,.\nonumber \\
 &  & \ \label{eq:QjcBc}
\end{eqnarray}
The correlation function on the right hand side of (\ref{eq:QjcBc})
can be evaluated by plugging
\begin{eqnarray}
\left[Q,\mathcal{G}(L,\Lambda,\delta)\right] & = & \int_{P_{L,\Lambda,\delta}}\frac{dz}{2\pi i}4\partial X^{0}\left(z\right)\bar{c}\bar{\partial}X^{0}\left(\bar{z}\right)-\int_{\bar{P}_{L,\Lambda,\delta}}\frac{d\bar{z}}{2\pi i}4\bar{\partial}X^{0}\left(\bar{z}\right)c\partial X^{0}\left(z\right)\nonumber \\
 &  & -2\left(c\partial X^{0}\left(i\infty\right)+\bar{c}\bar{\partial}X^{0}\left(-i\infty\right)\right)\left(\int_{P_{L,\Lambda,\delta}}\frac{dz}{2\pi i}\partial X^{0}\left(z\right)-\int_{\bar{P}_{L,\Lambda,\delta}}\frac{d\bar{z}}{2\pi i}\bar{\partial}X^{0}\left(\bar{z}\right)\right)\nonumber \\
 &  & +\int_{P_{L,\Lambda,\delta}}\frac{dz}{2\pi i}\frac{1}{2}\partial^{2}c-\int_{\bar{P}_{L,\Lambda,\delta}}\frac{d\bar{z}}{2\pi i}\frac{1}{2}\bar{\partial}^{2}\bar{c}\nonumber \\
 &  & +\int_{P_{L,\Lambda,\delta}}dz\partial\kappa\left(z,\bar{z}\right)+\int_{\bar{P}_{L,\Lambda,\delta}}d\bar{z}\bar{\partial}\kappa\left(z,\bar{z}\right)\,,\label{eq:contourintegral}\\
\kappa\left(z,\bar{z}\right) & \equiv & \frac{1}{\pi i}\left(X^{0}\left(z,\bar{z}\right)-X^{0}\left(i\infty,-i\infty\right)\right)\left(c\partial X^{0}\left(z\right)-\bar{c}\bar{\partial}X^{0}\left(\bar{z}\right)\right)\,,\nonumber 
\end{eqnarray}
into it and rewriting the result in terms of the operator formalism.
We need to take into account the fact that the correlation functions
are defined with time ordering with respect to the time variable $\mathrm{Re}z$. 

Since for $\mathrm{Im}z,\mathrm{\, Im}z^{\prime}\sim\infty$, 
\begin{eqnarray*}
\left\langle \partial X^{0}\left(z\right)\bar{\partial}X^{0}\left(\bar{z}^{\prime}\right)\right\rangle _{C_{L}} & \sim & -2\left(\frac{\pi}{L}\right)^{2}\exp\left(\frac{2\pi i}{L}\left(z-\bar{z}^{\prime}\right)\right)\,,\\
c\left(z\right) & \propto & \exp\left(-\frac{2\pi i}{L}z\right)\,,
\end{eqnarray*}
we can ignore the $\mathrm{Im}z=\Lambda$ part of the contours $P_{L,\Lambda,\delta},\bar{P}_{L,\Lambda,\delta}$
in the first and the second terms of (\ref{eq:contourintegral}),
in the limit $\Lambda\to\infty$. One can see that the contributions
from the terms on the second and the third lines of (\ref{eq:contourintegral})
vanish in the limit $\delta\to0$, because of the boundary conditions
of $X^{0},c,\bar{c}$. 

\begin{figure}
\begin{centering}
\includegraphics[scale=0.7]{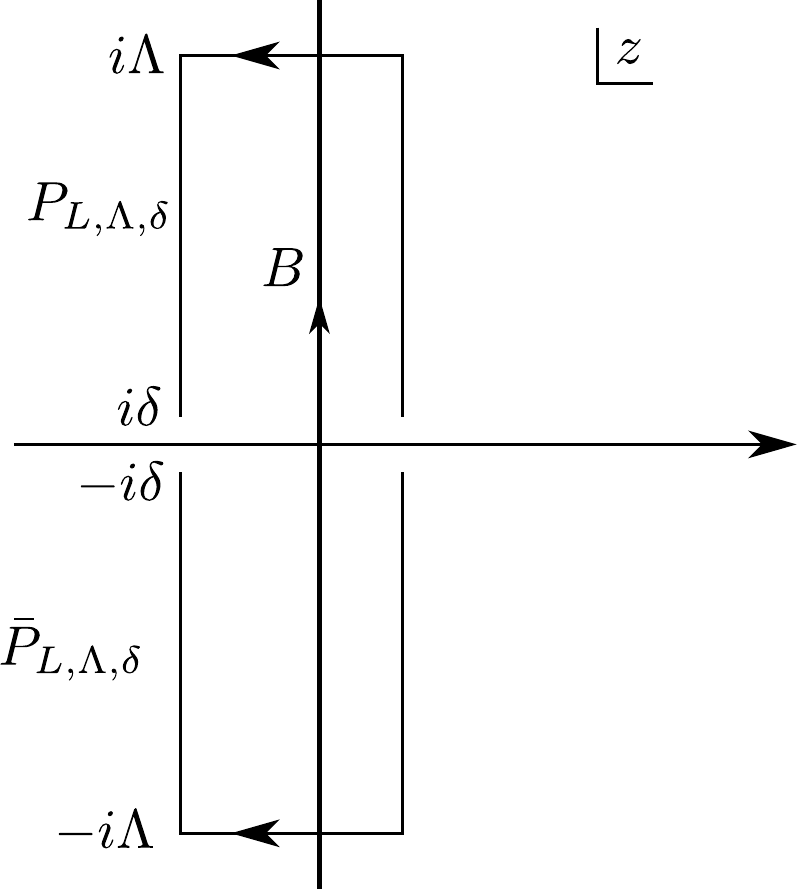}
\par\end{centering}

\caption{$P_{L,\Lambda,\delta}$ and $B$ in $\mathcal{A}_{1}$\label{fig:CLBA1}}
\end{figure}

In calculating the contribution of the terms on the fourth line of
(\ref{eq:contourintegral}), we need to be careful because the contours
$P_{L,\Lambda,\delta},\bar{P}_{L,\Lambda,\delta}$ intersect the contour
for $B=\int_{a-i\infty}^{a+i\infty}\frac{dz}{2\pi i}b$ as depicted
in Fig. \ref{fig:CLBA1}. We obtain
\begin{eqnarray*}
 &  & \left\langle e^{\left(L+\frac{1}{2}\right)K}f\circ\phi\left(0\right)e^{-\left(L+\frac{1}{2}\right)K}\left(\int_{P_{L,\Lambda,\delta}}dz\partial\kappa\left(z,\bar{z}\right)+\int_{\bar{P}_{L,\Lambda,\delta}}d\bar{z}\bar{\partial}\kappa\left(z,\bar{z}\right)\right)c\left(L_{2}+L_{3}\right)Bc\left(L_{3}\right)\right\rangle _{C_{L+1}}\\
 &  & \quad=-\mathrm{Tr}\left[e^{-\frac{1}{2}K}f\circ\phi\left(0\right)e^{-\left(L+\frac{1}{2}\right)K}c\left(L_{2}+L_{3}\right)Bc\left(L_{3}\right)\kappa\left(i\delta,-i\delta\right)\right.\\
 &  & \hphantom{\quad=\frac{1}{2\pi i}\mathrm{Tr}\quad}+e^{-\frac{1}{2}K}f\circ\phi\left(0\right)e^{-\left(L+\frac{1}{2}\right)K}\kappa\left(L_{1}+i\delta,L_{1}-i\delta\right)c\left(L_{2}+L_{3}\right)Bc\left(L_{3}\right)\\
 &  & \hphantom{\quad=\frac{1}{2\pi i}\mathrm{Tr}\quad}\left.+e^{-\frac{1}{2}K}f\circ\phi\left(0\right)e^{-\left(L+\frac{1}{2}\right)K}c\left(L_{2}+L_{3}\right)\left\{ B,\kappa\left(a+i\Lambda,a-i\Lambda\right)\right\} c\left(L_{3}\right)\right]\,.
\end{eqnarray*}
Putting all these pieces together and taking the limit $\Lambda\to\infty$,
we obtain
\begin{eqnarray}
\mathcal{A}_{1} & = & \int dL\mathrm{Tr}\left[e^{-\frac{1}{2}K}f\circ\phi\left(0\right)e^{-\frac{1}{2}K}\left(\chi e^{-LK}\psi\left(L\right)+e^{-LK}\psi\left(L\right)\chi\right)\right]\nonumber \\
 &  & +\int dL\frac{1}{L+1}\mathrm{Tr}\left[e^{-\frac{1}{2}K}f\circ\phi\left(0\right)e^{-\frac{1}{2}K}e^{-LK}\alpha\left(L\right)\right]\,,\label{eq:mathcalA12}
\end{eqnarray}
where $\alpha\left(L\right)$ is defined in (\ref{eq:alpha}) and
$\chi$ here is given as
\begin{eqnarray*}
\chi & = & \lim_{\left(\Lambda,\delta\right)\to\left(\infty,0\right)}\left[\int_{i\delta}^{i\Lambda}\frac{dz}{2\pi i}4\partial X^{0}\left(z\right)\bar{c}\bar{\partial}X^{0}\left(\bar{z}\right)\right.\\
 &  & \hphantom{\lim_{\left(\Lambda,\delta\right)\to\left(\infty,0\right)}\quad}-\int_{-i\delta}^{-i\Lambda}\frac{d\bar{z}}{2\pi i}4\bar{\partial}X^{0}\left(\bar{z}\right)c\partial X^{0}\left(z\right)\\
 &  & \hphantom{\lim_{\left(\Lambda,\delta\right)\to\left(\infty,0\right)}\quad}\left.\vphantom{\int_{i\delta}^{i\Lambda}\frac{dz}{2\pi i}4\partial X^{0}\left(z\right)\bar{c}\bar{\partial}X^{0}\left(\bar{z}\right)}+\frac{c\left(0\right)}{2\pi\delta}\right]\,.
\end{eqnarray*}

$\mathcal{A}_{2}$ is evaluated by substituting (\ref{eq:tildeQPsi2})
into the right hand side of (\ref{eq:mathcalA2}). Since $\mathcal{G}\left(0,\Lambda,\delta\right)=0$,
\begin{eqnarray}
\mathcal{A}_{2} & = & \lim_{\left(\Lambda,\delta\right)\to\left(\infty,0\right)}\int_{0}^{\infty}dL\left\langle e^{\left(L+\frac{1}{2}\right)K}f\circ\phi\left(0\right)e^{-\left(L+\frac{1}{2}\right)K}\right.\nonumber \\
 &  & \hphantom{\lim_{\left(\Lambda,\delta\right)\to\left(\infty,0\right)}\int_{0}^{\infty}dL\quad e^{\left(L+\frac{1}{2}\right)K}}\times\left.\vphantom{e^{\left(L+\frac{1}{2}\right)K}f\circ\phi\left(0\right)e^{-\left(L+\frac{1}{2}\right)K}}\mathcal{G}(L,\Lambda,\delta)e^{LK}\partial_{L}\left(e^{-LK}\alpha\left(L\right)\right)\right\rangle _{C_{L+1}}\,.\label{eq:mathcalA21}
\end{eqnarray}
The correlation function in the integrand can be rewritten as 
\[
\left.\partial_{t}\left\langle e^{\left(L+\frac{1}{2}\right)K}f\circ\phi\left(0\right)e^{-\left(L+\frac{1}{2}\right)K}\mathcal{G}(L,\Lambda,\delta)e^{-tK}\alpha\left(L+t\right)\right\rangle _{C_{L+1}}\right|_{t=0}\,,
\]
which can be evaluated in the limit $\Lambda\to\infty$, using (\ref{eq:X-X0CL})
as 
\begin{eqnarray*}
 &  & \left.\partial_{t}\left[\frac{L}{L+t+1}\left\langle e^{\left(L+\frac{1}{2}\right)K}f\circ\phi\left(0\right)e^{-\left(L+\frac{1}{2}\right)K}\alpha\left(L+t\right)\right\rangle _{C_{L+t+1}}\right]\right|_{t=0}\\
 &  & \quad=\frac{L}{L+1}\left\langle e^{\left(L+\frac{1}{2}\right)K}f\circ\phi\left(0\right)e^{-\left(L+\frac{1}{2}\right)K}e^{LK}\partial_{L}\left(e^{-LK}\alpha\left(L\right)\right)\right\rangle _{C_{L+1}}\\
 &  & \hphantom{\quad=\frac{L}{L+1}}-\frac{L}{\left(L+1\right)^{2}}\left\langle e^{\left(L+\frac{1}{2}\right)K}f\circ\phi\left(0\right)e^{-\left(L+\frac{1}{2}\right)K}\alpha\left(L\right)\right\rangle _{C_{L+1}}\,.
\end{eqnarray*}
Substituting this into (\ref{eq:mathcalA21}), we obtain
\begin{eqnarray*}
\mathcal{A}_{2} & = & \int_{0}^{\infty}dL\left\{ \frac{L}{L+1}\mathrm{Tr}\left[e^{-\frac{1}{2}K}f\circ\phi\left(0\right)e^{-\frac{1}{2}K}\partial_{L}\left(e^{-LK}\alpha\left(L\right)\right)\right]\right.\\
 &  & \hphantom{\int_{0}^{\infty}dL\quad}\left.-\frac{L}{\left(L+1\right)^{2}}\mathrm{Tr}\left[e^{-\frac{1}{2}K}f\circ\phi\left(0\right)e^{-\frac{1}{2}K}e^{-LK}\alpha\left(L\right)\right]\right\} \\
 & = & -\int_{0}^{\infty}dL\frac{1}{L+1}\mathrm{Tr}\left[e^{-\frac{1}{2}K}f\circ\phi\left(0\right)e^{-\frac{1}{2}K}e^{-LK}\alpha\left(L\right)\right]\,.
\end{eqnarray*}

Putting these together, we get 
\[
\mathcal{A}_{1}+\mathcal{A}_{2}=\int dL\mathrm{Tr}\left[e^{-\frac{1}{2}K}f\circ\phi\left(0\right)e^{-\frac{1}{2}K}\left(\chi e^{-LK}\psi\left(L\right)+e^{-LK}\psi\left(L\right)\chi\right)\right]\,.
\]
The right hand side coincides with $\left\langle \phi\right|\left(\chi-\chi^{\dagger}\right)\left|\Psi\right\rangle $
and thus we obtain
\[
\left[Q,\mathcal{G}\right]\left|\Psi\right\rangle =\left(\chi-\chi^{\dagger}\right)\left|\Psi\right\rangle \,,
\]
 in the setup in this section. %

\subsection{(\ref{eq:Egio}) for Okawa type solutions}

With (\ref{eq:QmathcalGPsi}) established, it is straightforward to
follow the procedure given in (\ref{eq:Egio2}) and prove (\ref{eq:Egio}).
In summary, we have proved (\ref{eq:Egio}) for Okawa type solutions
$\Psi$ assuming the following conditions:
\begin{itemize}
\item $\Psi$ satisfies the equation of motion.
\item $\alpha\left(\infty\right)=0$ and $\alpha\left(0\right)$ is well-defined
for $\alpha\left(L\right)$ defined in (\ref{eq:alpha}). 
\end{itemize}
In addition to these, it is implicitly assumed that all the quantities
which appear in the course of the calculations are finite%
\footnote{This is also assumed in section \ref{sec:derivation}.%
}. Conditions other than the equation of motion are concerning the
regularity of the solution. If the equation of motion is not satisfied,
we obtain (\ref{eq:anomaly}) with $\left|\Gamma\right\rangle $ given
in (\ref{eq:Gamma}).

\section{Other solutions\label{sec:Other-solutions}}

We can use the method in the previous section and prove (\ref{eq:Egio})
for other types of solutions%
\footnote{Our results will not be useful for the marginal deformation solutions,
for which it is trivial to calculate the energy, but may be relevant
\cite{Okawa2012} in the context of the discussions in Ref.\ \cite{Kudrna:2012um}. %
}. We will discuss BMT solution and Murata-Schnabl solution in the
following.

\subsection{BMT solution}

In \cite{Bonora:2010hi}, Bonora, Maccaferri, and Tolla construct
solutions corresponding to relevant deformations of BCFT , called
BMT solution%
\footnote{An earlier proposal for such solutions were made in \cite{Ellwood:2009zf}%
}. They enlarged the $K,B,c$ algebra by adding a relevant matter operator
$\phi$ which satisfies 
\begin{eqnarray*}
\lim_{s\to0}s\phi\left(s\right)\phi\left(0\right) & = & 0\,,\\
\left[c,\phi\right]=\left[B,\phi\right] & = & 0\,,\\
Q\phi & = & c\partial\phi+\partial c\delta\phi\,.
\end{eqnarray*}
The BMT solution is given as 
\begin{equation}
\Psi=c\phi-\frac{1}{K+\phi}\left(\phi-\delta\phi\right)Bc\partial c\,.\label{eq:BMT}
\end{equation}
In order to realize the lump solution, $\phi$ is usually taken to
be the so-called Witten deformation 
\[
\phi\left(s\right)=u\left(\frac{1}{2}:X^{2}:\left(s\right)+\gamma-1+\ln\left(2\pi u\right)\right)\,,
\]
or the cosine deformation 
\[
\phi(s)=u\left[-u^{-1/R^{2}}:\cos\left(\frac{1}{R}X\right):(s)+A(R)\right]\,.
\]
Here $X$ direction is noncompact for Witten deformation and a circle
of radius $R>\sqrt{2}$ for the cosine deformation. $A\left(R\right)$
is a constant determined in \cite{Bonora:2010hi}. 

If one tries to define the $\frac{1}{K+\phi}$ which appears in the
BMT solution as 
\[
\frac{1}{K+\phi}\equiv\int_{0}^{\infty}dte^{-t\left(K+\phi\right)}\,,
\]
via the Schwinger parametrization, the integral on the right hand
side diverges because $\lim_{t\to\infty}e^{-t\left(K+\phi\right)}$
coincides with the deformed sliver state $\tilde{\Omega}^{\infty}$.
One way to regularize the divergence is to replace $\frac{1}{K+\Phi}$
by $\frac{1}{K+\phi+\epsilon}$ with $1\gg\epsilon>0$ and consider
\[
\Psi_{\epsilon}=c\phi-\frac{1}{K+\phi+\epsilon}\left(\phi-\delta\phi\right)Bc\partial c\,,
\]
but $\Psi_{\epsilon}$ suffers from an anomaly in equation of motion
\cite{Erler:2011tc}:
\[
Q\Psi_{\epsilon}+\Psi_{\epsilon}^{2}=\Gamma_{\epsilon}\equiv\frac{\epsilon}{K+\phi+\epsilon}\left(\phi-\delta\phi\right)c\partial c\,.
\]
In \cite{Bonora:2011ns}, the authors propose a way to deal with the
problem using the distribution theory. 

It is quite easy to compute the gauge invariant observables for the
BMT solution, but it is much more difficult to calculate the energy.
Our method can be used to improve the situation a bit. In \cite{Bonora:2011ri,Erler:2011tc},
the authors define a solution 
\[
\Psi_{\epsilon}^{\prime}=c\left(\phi+\epsilon\right)-\frac{1}{K+\phi+\epsilon}\left(\phi-\delta\phi+\epsilon\right)Bc\partial c\,,
\]
as a possible regularization of the BMT solution, but it actually
describes the tachyon vacuum. It is shown that if the energy of the
solution $\Psi_{\epsilon}^{\prime}$ is that of the tachyon vacuum,
one can prove analytically that the energy of the BMT solution coincides
with that of the lump solution \cite{Erler:2011tc,Bonora:2011ri}.
The gauge invariant observables of $\Psi_{\epsilon}^{\prime}$ can
be calculated analytically, which turn out to be equal to those of
the tachyon vacuum but the energy is calculated only numerically \cite{Bonora:2011ri,Bonora:2011ru}
in the case of the Witten deformation. We will use our method to calculate
the energy of $\Psi_{\epsilon}^{\prime}$. It is quite straightforward
to generalize the calculations in the previous section to $\Psi_{\epsilon}^{\prime}$,
starting from the Laplace transformed form 
\begin{eqnarray*}
\Psi_{\epsilon}^{\prime} & = & \int_{0}^{\infty}dLe^{-LK}\psi_{\epsilon}^{\prime}\left(L\right)\,,\\
\psi_{\epsilon}^{\prime}\left(L\right) & = & \delta\left(L\right)c\left(\phi+\epsilon\right)-e^{-\epsilon L-\int_{0}^{L}ds\phi\left(s\right)}\left(\phi-\delta\phi+\epsilon\right)Bc\partial c\,,
\end{eqnarray*}
where 
\[
\phi\left(s\right)=e^{sK}\phi e^{-sK}\,,
\]
and the operators are time ordered. For $\Psi_{\epsilon}^{\prime}$,
one can obtain 
\[
\mathcal{L}^{-1}\left\{ Q\Psi_{\epsilon}^{\prime}\right\} \left(L\right)=Q\mathcal{L}^{-1}\left\{ \Psi_{\epsilon}^{\prime}\right\} \left(L\right)-e^{LK}\partial_{L}\left(e^{-LK}\alpha_{\epsilon}^{\prime}\left(L\right)\right)-\delta\left(L\right)\alpha_{\epsilon}^{\prime}\left(0\right)\,,
\]
with
\[
\alpha_{\epsilon}^{\prime}\left(L\right)=e^{-\epsilon L-\int_{0}^{L}ds\phi\left(s\right)}\left(\phi-\delta\phi\right)c\partial c\,.
\]
Since $\alpha_{\epsilon}^{\prime}\left(\infty\right)=0$ and $\alpha_{\epsilon}^{\prime}\left(0\right)$
is well-defined, all the manipulations in the previous section are
valid provided that we do not encounter any divergences in the course
of the calculations. In the case of the Witten deformation, there
exist divergences coming from noncompactness of the direction corresponding
to $X$ and our method is not applicable. $\Psi_{\epsilon}^{\prime}$
corresponding to the cosine deformation does not seem to have such
a problem%
\footnote{The partition function 
\[
g\left(uT\right)\equiv\mathrm{Tr}e^{-T\left(K+\phi\right)}\,,
\]
can be calculated perturbatively \cite{Fendley:1994ms} and is finite
for $0\leq uT<\infty$. The UV and IR behaviors of the correlation
functions of $\phi$'s are harmless. %
} and we can see that the energy coincides with that of the tachyon
vacuum. 

It may be possible to calculate the energy of $\Psi_{\epsilon}$ directly
for the cosine deformation. Since $\Psi_{\epsilon}$ has an anomaly
in equation of motion, we need to evaluate the second and the third
terms of (\ref{eq:anomaly}). In order to do so, we need to know the
IR behavior of some correlation functions of $\phi$. 

\subsection{Murata-Schnabl solution}

Murata and Schnabl \cite{Murata:2011ex,Murata:2011ep} propose that
the Okawa type solution (\ref{eq:Okawa}) with
\begin{eqnarray}
G\left(K\right) & \equiv & 1-F^{2}\left(K\right)\nonumber \\
 & = & \left(\frac{K+1}{K}\right)^{N-1}\,,\label{eq:MurataSchnabl}
\end{eqnarray}
corresponds to a configuration with $N$ D-branes. Since the solution
itself is singular for $N\ne0,1$, the authors need some regularization
in calculating various quantities. In order to define the gauge invariant
observables, they replace $K$ by $K+\epsilon\,\left(\epsilon\ll1\right)$
and consider
\[
F\left(K+\epsilon\right)c\frac{B\left(K+\epsilon\right)}{1-F^{2}\left(K+\epsilon\right)}cF\left(K+\epsilon\right)\,.
\]
or its gauge equivalent 
\[
\Psi_{\epsilon}=F^{2}\left(K+\epsilon\right)cB\frac{K+\epsilon}{1-F^{2}\left(K+\epsilon\right)}c\,.
\]
The energy of the solution is calculated using a different way to
regularize the divergence. They obtain the energy and the gauge invariant
observables which coincide with those for $N$ D-branes. It is necessary
to find a more solid way to define the solution, and there are many
attempts to rectify the situation \cite{Takahashi:2011wk,Hata:2011ke,Erler:2012qn,Erler:2012qr,Masuda:2012kt}. 

As an application of our results, let us calculate the energy of $\Psi_{\epsilon}$
in this paper. Since $\Psi_{\epsilon}$ has an anomaly in equation
of motion, 
\[
Q\Psi_{\epsilon}+\Psi_{\epsilon}^{2}=\Gamma_{\epsilon}\,,
\]
where
\begin{eqnarray*}
\Gamma_{\epsilon} & = & \epsilon\left(1-G_{\epsilon}\left(K\right)\right)c\frac{K+\epsilon}{G_{\epsilon}\left(K\right)}c\,,\\
G_{\epsilon}\left(K\right) & \equiv & G\left(K+\epsilon\right)\,,
\end{eqnarray*}
the relation we have is
\begin{equation}
E=\frac{1}{g^{2}}\left[\left\langle I|\mathcal{V}\left(i\right)|\Psi_{\epsilon}\right\rangle -\left\langle I\right|\chi\left|\Gamma_{\epsilon}\right\rangle +\left\langle \mathcal{G}\Psi_{\epsilon}|\Gamma_{\epsilon}\right\rangle \right]\,,\label{eq:anomaly2}
\end{equation}
which can be proved as in the previous section. After some calculations,
details of which are presented in appendix \ref{sec:Calculation-of-the},
we obtain in the limit $\epsilon\to0$
\begin{eqnarray}
\left\langle I|\mathcal{V}\left(i\right)|\Psi_{\epsilon}\right\rangle  & = & \frac{N-1}{2\pi^{2}}\nonumber \\
\left\langle I\right|\chi\left|\Gamma_{\epsilon}\right\rangle  & \to & R_{N}\,,\label{eq:RN}\\
\left\langle \mathcal{G}\Psi_{\epsilon}|\Gamma_{\epsilon}\right\rangle  & \to & 0\,,\label{eq:GPsiGamma}
\end{eqnarray}
where 
\begin{eqnarray*}
R_{N} & \equiv & \begin{cases}
-\frac{i}{8\pi^{3}}\sum_{k=0}^{N-2}\frac{N!}{k!\left(k+2\right)!\left(N-2-k\right)!}\left((2\pi i)^{k+2}-\left(-2\pi i\right)^{k+2}\right) & ,\ \left(N\geq1\right)\,,\\
\frac{i}{8\pi^{3}}\sum_{k=0}^{-N-1}\frac{\left(1-N\right)!}{k!\left(k+2\right)!\left(-N-1-k\right)!}\left((2\pi i)^{k+2}-\left(-2\pi i\right)^{k+2}\right) & ,\ \left(N\leq0\right)\,.
\end{cases}
\end{eqnarray*}
Therefore we get the energy
\[
E=\frac{1}{g^{2}}\left(\frac{N-1}{2\pi^{2}}-R_{N}\right)\,.
\]
This coincides with the desired value $\frac{N-1}{2\pi^{2}}$ for
$N=-1,0,1,2$. Thus, for these $N$, the anomaly $\Gamma_{\epsilon}$
is harmless at least in the calculation of energy, although we do
not know the reason why this is so for $N=-1,2$ %
\footnote{$N=-1,2$ may be argued to be special in the following sense. $\Psi_{\varepsilon}$
is gauge equivalent to
\[
\epsilon\left(\frac{1}{G_{\epsilon}}-1\right)cBG_{\epsilon}c\,,
\]
which is regular in the limit $\epsilon\to0$ for $N=-1,2$. Another
reason for $N=-1$ may be because there exists a regular solution
\cite{Masuda:2012kt}. %
}.

\section{Conclusion and discussion\label{sec:Conclusion-and-discussion}}

In this paper, we present a way to show that the energy is proportional
to a gauge invariant observable, which corresponds to the graviton
one point function, for a classical solution in Witten's cubic open
string field theory. We give a method which can be used to show this
even for the solutions which involve $K,B$. Usually the gauge invariant
observables are much easier to calculate compared with the energy.
In a recent paper \cite{Kudrna:2012re}, it is found that the boundary
states can also be constructed from the gauge invariant observables.
Therefore now we possess a more efficient way to study the physical
properties of solutions which have been or will be discovered. 

Recently in \cite{Masuda:2012kt} the authors propose several new
types of solutions made from $K,B,c$. It seems that our method can
be applied to these solutions and derive (\ref{eq:Egio}) if the solutions
are sufficiently regular. One particularly interesting solution mentioned
in \cite{Masuda:2012kt} is the one due to Masuda, which is claimed
to have the energy of the double brane configuration but the gauge
invariant observables of the perturbative vacuum. It would be intriguing
to check how our derivation of (\ref{eq:Egio}) fails for this solution. 

Interrelationship between energy and the gauge invariant observable
will be important in exploring various aspects of string fields. For
example, in the case of the BMT solution, the calculation of gauge
invariant observables reduces to the integral of total derivative.
This implies that these gauge invariant observables may have some
topological nature. On the other hand, in \cite{Hata:2011ke}, the
energy is interpreted to be the winding number in string field theory.
Our results may shed some light on the study of the topological invariants
of the space of string fields.

\section*{Acknowledgments}

We would like to thank K. Murakami and Y. Satoh for discussions. N.
I. would like to thank M. Schnabl, Y. Okawa and other participants
of SFT 2012 for questions, suggestions and correspondence. This work
was supported in part by Grant-in-Aid for Scientific Research (C)
(20540247) from MEXT. 

\appendix

\section{Derivations of (\ref{eq:Qchi}), (\ref{eq:ShQprime}) and (\ref{eq:QmathcalG})
\label{sec:Derivation-of-}}

Since the quantities which appear in section \ref{sec:derivation}
involve unusual combinations of operators, some explanation is necessary
about the definitions and the treatment of them. In this appendix,
we present the details of the definition of $\chi,\mathcal{G}$ and
the derivation of (\ref{eq:Qchi})(\ref{eq:ShQprime})(\ref{eq:QmathcalG}).

\subsection*{$\left\{ Q,\chi\right\} =\mathcal{V}\left(i,-i\right)$}

Introducing $\theta$ such that $\xi=e^{i\theta}$, the contour integral
on the right hand side of (\ref{eq:chi}) is expressed as 
\begin{eqnarray}
 &  & \int_{P_{1}}\frac{d\xi}{2\pi i}j\left(\xi,\bar{\xi}\right)-\int_{\bar{P}_{1}}\frac{d\bar{\xi}}{2\pi i}\bar{j}\left(\xi,\bar{\xi}\right)\nonumber \\
 &  & \qquad=\int_{\delta}^{\frac{\pi}{2}}\frac{d\theta}{2\pi i}ie^{i\theta}j\left(e^{i\theta},e^{-i\theta}\right)-\int_{\delta}^{\frac{\pi}{2}}\frac{d\theta}{2\pi i}\left(-ie^{-i\theta}\right)\bar{j}\left(e^{i\theta},e^{-i\theta}\right)\,.\label{eq:inttheta}
\end{eqnarray}
In calculating the BRST variation of this quantity, it is useful to
notice
\begin{eqnarray}
\frac{1}{2\pi i}j\left(\xi,\bar{\xi}\right) & = & \oint_{\xi}\frac{d\xi^{\prime}}{2\pi i}b\left(\xi^{\prime}\right)\mathcal{V}\left(\xi,\bar{\xi}\right)\,,\label{eq:j}\\
-\frac{1}{2\pi i}\bar{j}\left(\xi,\bar{\xi}\right) & = & \oint_{\bar{\xi}}\frac{d\bar{\xi}^{\prime}}{2\pi i}\bar{b}\left(\bar{\xi}^{\prime}\right)\mathcal{V}\left(\xi,\bar{\xi}\right)\,,\label{eq:jbar}
\end{eqnarray}
where $\mathcal{V}\left(\xi,\bar{\xi}\right)$ is the vertex operator
defined in (\ref{eq:constantgraviton}). Since $\mathcal{V}$ is BRST
invariant, it is straightforward to show
\begin{eqnarray}
 &  & \left\{ Q,\int_{P_{1}}\frac{d\xi}{2\pi i}j\left(\xi,\bar{\xi}\right)-\int_{\bar{P}_{1}}\frac{d\bar{\xi}}{2\pi i}\bar{j}\left(\xi,\bar{\xi}\right)\right\} \nonumber \\
 &  & \quad=\int_{\delta}^{\frac{\pi}{2}}d\theta\left(\frac{de^{i\theta}}{d\theta}\partial_{\xi}\mathcal{V}\left(e^{i\theta},e^{-i\theta}\right)+\frac{de^{-i\theta}}{d\theta}\partial_{\bar{\xi}}\mathcal{V}\left(e^{i\theta},e^{-i\theta}\right)\right)\nonumber \\
 &  & \quad=\mathcal{V}\left(i,-i\right)-\mathcal{V}\left(e^{i\delta},e^{-i\delta}\right)\,.\label{eq:Vi-Vdelta}
\end{eqnarray}
Assuming that there are no other operators around $\xi=1$, the OPE's
of $c,\bar{c},X^{0}$ imply 
\begin{equation}
\mathcal{V}\left(e^{i\delta},e^{-i\delta}\right)=\frac{c\partial c\left(1\right)}{2\pi\delta}+\mathcal{O}\left(\delta\right)=\left\{ Q,\frac{c\left(1\right)}{2\pi\delta}\right\} +\mathcal{O}\left(\delta\right)\,,\label{eq:Veidelta}
\end{equation}
for $\delta\sim0$. The assumption is valid in the setup of this paper.
Using (\ref{eq:Veidelta}), we obtain
\[
\left\{ Q,\chi\right\} =\mathcal{V}\left(i,-i\right)\,.
\]

It is possible to generalize our construction here to other closed
string vertex operators. For any BRST invariant closed string vertex
operator $\mathcal{V}\left(\xi,\bar{\xi}\right)$, one can define
$j,\bar{j}$ as in (\ref{eq:j})(\ref{eq:jbar}), and one can prove
(\ref{eq:Vi-Vdelta}). If $\mathcal{V}\left(e^{i\delta},e^{-i\delta}\right)$
can be expressed as
\begin{equation}
\mathcal{V}\left(e^{i\delta},e^{-i\delta}\right)=\left\{ Q,\mathcal{U}\right\} +\mathcal{O}\left(\delta\right)\,,\label{eq:Veidelta1}
\end{equation}
in the limit $\delta\to0$ as in (\ref{eq:Veidelta}), we obtain $\mathcal{V}\left(i,-i\right)=\left\{ Q,\chi\right\} $
with 
\[
\chi\equiv\lim_{\delta\to0}\left[\int_{P_{1}}\frac{d\xi}{2\pi i}j\left(\xi,\bar{\xi}\right)-\int_{\bar{P}_{1}}\frac{d\bar{\xi}}{2\pi i}\bar{j}\left(\xi,\bar{\xi}\right)+\mathcal{U}\right]\,.
\]
(\ref{eq:Veidelta1}) holds if there exists no on-shell open string
vertex operator $V_{o}$ such that 
\[
\left\langle \mathcal{V}V_{o}\right\rangle _{\mathrm{disk}}\ne0\,.
\]

\subsection*{(\ref{eq:ShQprime})}

\begin{figure}[h]
\begin{centering}
\includegraphics[scale=0.5]{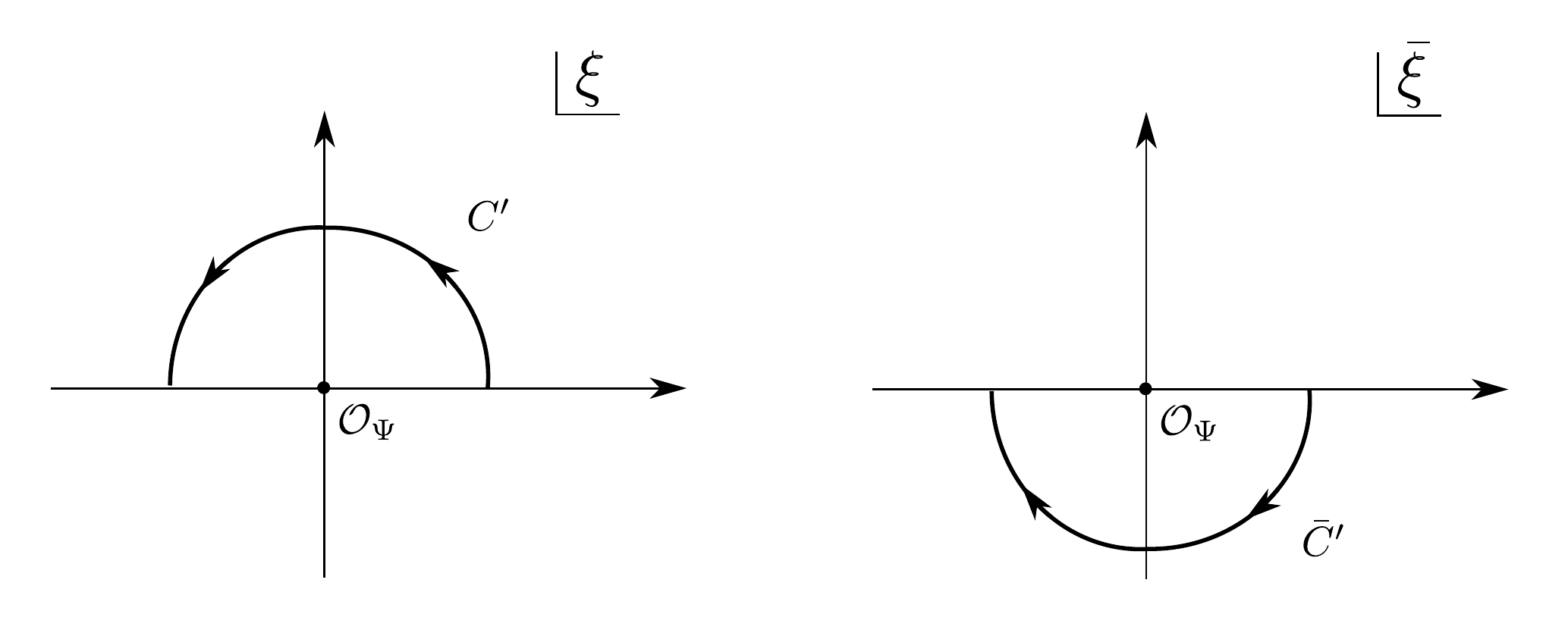}
\par\end{centering}

\caption{$C^{\prime}$\label{fig:Cprime}}
\end{figure}

\begin{figure}[h]
\begin{centering}
\includegraphics[scale=0.6]{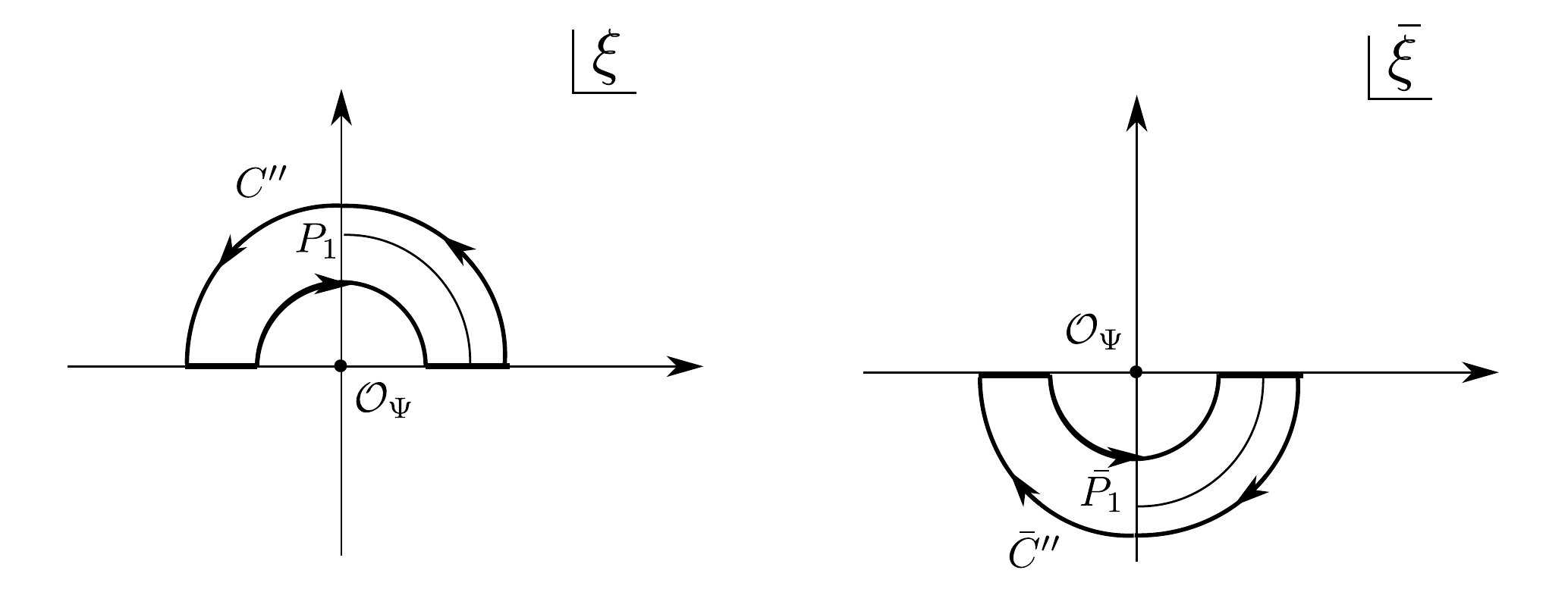}
\par\end{centering}

\caption{Contour which surrounds $P_{1}$\label{fig:Contour-which-surrounds}}

\end{figure}

Substituting (\ref{eq:Psiprime}) into (\ref{eq:Sh}), we obtain
\begin{eqnarray*}
S_{h} & = & -\frac{1}{g^{2}}\left[\frac{1}{2}\left\langle \Psi^{\prime}|Q|\Psi^{\prime}\right\rangle +\frac{1}{3}\left\langle \Psi^{\prime}|\Psi^{\prime}*\Psi^{\prime}\right\rangle +h\left\langle I|\mathcal{V}\left(i\right)|\Psi\right\rangle \right.\\
 &  & \hphantom{-\frac{1}{g^{2}}\quad}\left.-h\left\langle I\right|\chi Q\left|\Psi\right\rangle -\frac{h}{2}\left\langle \Psi^{\prime}\right|\left(\chi-\chi^{\dagger}\right)\left|\Psi^{\prime}\right\rangle \right]\,,
\end{eqnarray*}
where we have used 
\begin{eqnarray*}
\chi\left|I\right\rangle  & = & \chi^{\dagger}\left|I\right\rangle \,,\\
\left\langle \Psi\right|\chi\left|\Psi\right\rangle  & = & -\left\langle \Psi\right|\chi^{\dagger}\left|\Psi\right\rangle \,.
\end{eqnarray*}
Since $Q\left|I\right\rangle =0$, 
\[
\left\langle I\right|\chi Q\left|\Psi\right\rangle =\left\langle I\right|\left\{ Q,\chi\right\} \left|\Psi\right\rangle \,,
\]
and we may be able to use (\ref{eq:Qchi}) to show (\ref{eq:ShQprime}).
We should check if the $Q$ in the open string field action yields
the BRST variation of $\chi$ as an operator in the bulk. The BRST
operator acting on a string field $\left|\Psi\right\rangle =\mathcal{O}_{\Psi}\left(0\right)\left|0\right\rangle $
is given as 
\begin{eqnarray*}
Q\left|\Psi\right\rangle  & = & \left(\int_{C^{\prime}}\frac{d\xi}{2\pi i}J_{\mathrm{B}}-\int_{\bar{C}^{\prime}}\frac{d\bar{\xi}}{2\pi i}\bar{J}_{\mathrm{B}}\right)\mathcal{O}_{\Psi}\left(0\right)\left|0\right\rangle \,,
\end{eqnarray*}
where $J_{\mathrm{B}},\bar{J}_{\mathrm{B}}$ are the BRST current
and $C^{\prime},\bar{C}^{\prime}$ are depicted in the figure \ref{fig:Cprime}.
Since $J_{\mathrm{B}}\left(\xi\right)=\bar{J}_{\mathrm{B}}\left(\bar{\xi}\right)$
for real $\xi$ the contour integral can be expressed as 
\[
\oint_{0}\frac{d\xi}{2\pi i}J_{\mathrm{B}}\,,
\]
on the doubled Riemann surface. $\left(Q\chi\left(i,-i\right)+\chi\left(i,-i\right)Q\right)\left|\Psi\right\rangle $
in the open string field theory is given as 
\[
\left(\oint_{C^{\prime\prime}}\frac{d\xi}{2\pi i}J_{\mathrm{B}}-\oint_{\bar{C}^{\prime\prime}}\frac{d\bar{\xi}}{2\pi i}\bar{J}_{\mathrm{B}}\right)\chi\left(\xi,\bar{\xi}\right)\mathcal{O}_{\psi}\left|0\right\rangle \,,
\]
where the contours $C^{\prime\prime},\bar{C}^{\prime\prime}$ are
the one which surrounds $P_{1}\bar{P}_{1}$ as depicted in figure
\ref{fig:Contour-which-surrounds}. Hence the contour integral yields
the BRST variation of $\chi$ and we obtain $\mathcal{V}\left(i,-i\right)\left|\Psi\right\rangle $.

\subsection*{$\left\{ Q,\mathcal{G}\right\} =\chi-\chi^{\dagger}$}

The contour integral on the right hand side of (\ref{eq:mathcalG})
is defined in the same way as in (\ref{eq:inttheta}). It is straightforward
to calculate the BRST variations of $g_{\xi},g_{\bar{\xi}}$ as 
\begin{eqnarray*}
\left[Q,g_{\xi}\left(\xi,\bar{\xi}\right)\right] & = & \frac{1}{2}\partial^{2}c\left(\xi\right)+\partial_{\xi}\left(2\left(X^{0}\left(\xi,\bar{\xi}\right)-X^{0}\left(i,-i\right)\right)c\partial X^{0}\left(\xi\right)\right)\\
 &  & +2\bar{c}\bar{\partial}X^{0}\partial X^{0}\left(\xi,\bar{\xi}\right)-2\left(c\partial X^{0}\left(i\right)+\bar{c}\bar{\partial}X^{0}\left(-i\right)\right)\partial X^{0}\left(\xi\right)\,,\\
\left[Q,g_{\bar{\xi}}\left(\xi,\bar{\xi}\right)\right] & = & \frac{1}{2}\bar{\partial}^{2}\bar{c}\left(\bar{\xi}\right)+\partial_{\bar{\xi}}\left(2\left(X^{0}\left(\xi,\bar{\xi}\right)-X^{0}\left(i,-i\right)\right)\bar{c}\bar{\partial}X^{0}\left(\bar{\xi}\right)\right)\\
 &  & +2c\partial X^{0}\bar{\partial}X^{0}\left(\xi,\bar{\xi}\right)-2\left(c\partial X^{0}\left(i\right)+\bar{c}\bar{\partial}X^{0}\left(-i\right)\right)\bar{\partial}X^{0}\left(\bar{\xi}\right)\,,
\end{eqnarray*}
and we find $\left[Q,\mathcal{G}\right]$ is equal to
\begin{eqnarray*}
 &  & \lim_{\delta\to0}\left[\frac{1}{4\pi i}\left(\partial c\left(-e^{-i\delta}\right)-\bar{\partial}c\left(-e^{i\delta}\right)-\partial c\left(e^{i\delta}\right)+\bar{\partial}\bar{c}\left(e^{-i\delta}\right)\right)\right.\\
 &  & \hphantom{\lim_{\delta\to0}\quad}+\frac{1}{2\pi i}\left(\int d\xi\partial_{\xi}+\int d\bar{\xi}\partial_{\bar{\xi}}\right)\left(2\left(X^{0}\left(\xi,\bar{\xi}\right)-X^{0}\left(i,-i\right)\right)c\partial X^{0}\left(\xi\right)\right)\\
 &  & \hphantom{\lim_{\delta\to0}\quad}-\frac{1}{2\pi i}\left(\int d\xi\partial_{\xi}+\int d\bar{\xi}\partial_{\bar{\xi}}\right)\left(2\left(X^{0}\left(\xi,\bar{\xi}\right)-X^{0}\left(i,-i\right)\right)\bar{c}\bar{\partial}X^{0}\left(\bar{\xi}\right)\right)\\
 &  & \hphantom{\lim_{\delta\to0}\quad}+\int_{P_{1}+P_{2}}\frac{d\xi}{2\pi i}4\partial X^{0}\bar{c}\bar{\partial}X^{0}\left(\xi,\bar{\xi}\right)-\int_{\bar{P}_{1}+\bar{P}_{2}}\frac{d\bar{\xi}}{2\pi i}4\bar{\partial}X^{0}c\partial X^{0}\left(\xi,\bar{\xi}\right)\\
 &  & \left.\hphantom{\lim_{\delta\to0}\quad}-2\left(c\partial X^{0}\left(i\right)+\bar{c}\bar{\partial}X^{0}\left(-i\right)\right)\left(\int_{P_{1}+P_{2}}\frac{d\xi}{2\pi i}\partial X^{0}\left(\xi\right)-\int_{\bar{P}_{1}+\bar{P}_{2}}\frac{d\bar{\xi}}{2\pi i}\bar{\partial}X^{0}\left(\bar{\xi}\right)\right)\right]\,.
\end{eqnarray*}
The terms on the first line cancel with each other in the limit $\delta\to0$
because of the boundary conditions of $c,\bar{c}$. Those on the fifth
vanish if $\mathcal{O}_{\Psi}$ does not involve $X^{0}$. The second
and the third lines yield in the limit $\delta\to0$
\begin{eqnarray*}
 &  & \left.\frac{1}{\pi i}\left(X^{0}\left(\xi,\bar{\xi}\right)-X^{0}\left(i,-i\right)\right)\left(c\partial X^{0}\left(\xi\right)-\bar{c}\bar{\partial}X^{0}\left(\bar{\xi}\right)\right)\right|_{\left(\xi,\bar{\xi}\right)=\left(e^{i\delta},e^{-i\delta}\right)}^{\left(-e^{-i\delta},-e^{i\delta}\right)}\\
 &  & \quad\sim-\frac{c\left(-1\right)}{2\pi\delta}+\frac{c\left(1\right)}{2\pi\delta}\,.
\end{eqnarray*}
Thus we get 
\[
\left[Q,\mathcal{G}\right]=\chi-\chi^{\dagger}\,.
\]

\section{Laplace transformed form of the string field\label{sec:Laplace-transformed-form}}

We derive two formulas (\ref{eq:noncommconv}) (\ref{eq:tildeQPsi2})
concerning the Laplace transform of the string field defined in section
\ref{sec:KBc}. 

For two string fields $A_{1},A_{2},$ which can be expressed as a
sum of wedge states with insertions, it is easy to show
\begin{equation}
\mathcal{L}^{-1}\left\{ A_{1}A_{2}\right\} \left(L\right)=\int_{0}^{L}dL^{\prime}e^{L^{\prime}K}\mathcal{L}^{-1}\left\{ A_{1}\right\} \left(L-L^{\prime}\right)e^{-L^{\prime}K}\mathcal{L}^{-1}\left\{ A_{2}\right\} \left(L^{\prime}\right)\,.\label{eq:noncommconv}
\end{equation}
The right hand side can be regarded as an operator version of convolution. 

For $\psi\left(L\right)$ in (\ref{eq:Psitilde}),
\begin{eqnarray}
Q\psi\left(L\right) & = & Q\mathcal{L}^{-1}\left\{ \Psi\right\} \left(L\right)\nonumber \\
 & = & \int dL_{1}dL_{2}dL_{3}\delta\left(L-L_{1}-L_{2}-L_{3}\right)\nonumber \\
 &  & \quad\times\left[c\partial c\left(L_{2}+L_{3}\right)Bc\left(L_{3}\right)-c\left(L_{2}+L_{3}\right)Kc\left(L_{3}\right)+c\left(L_{2}+L_{3}\right)Bc\partial c\left(L_{3}\right)\right]\nonumber \\
 &  & \quad\times f\left(L_{1}\right)\tilde{f}\left(L_{2}\right)f\left(L_{3}\right)\,,\label{eq:QtildePsi}
\end{eqnarray}
which is not equal to 
\begin{eqnarray}
\mathcal{L}^{-1}\left\{ Q\Psi\right\} \left(L\right) & = & \int dL_{1}dL_{2}dL_{3}\delta\left(L-L_{1}-L_{2}-L_{3}\right)\nonumber \\
 &  & \quad\times\left[\vphantom{\widetilde{\frac{K^{2}}{1-F^{2}}}\left(L_{2}\right)}\left\{ \partial c\left(L_{2}+L_{3}\right)Bc\left(L_{3}\right)+c\left(L_{2}+L_{3}\right)Bc\partial c\left(L_{3}\right)\right\} \right.\nonumber \\
 &  & \hphantom{\quad\times\qquad}\times f\left(L_{1}\right)\tilde{f}\left(L_{2}\right)f\left(L_{3}\right)\nonumber \\
 &  & \hphantom{\quad\times\quad}\left.-c\left(L_{2}+L_{3}\right)c\left(L_{3}\right)f\left(L_{1}\right)\mathcal{L}^{-1}\left\{ \frac{K^{2}}{1-F^{2}}\right\} \left(L_{2}\right)f\left(L_{3}\right)\right]\,.\label{eq:L-1QPsi}
\end{eqnarray}
Therefore the BRST transformation and $\mathcal{L}^{-1}$ do not commute
with each other. Comparing (\ref{eq:QtildePsi}) and (\ref{eq:L-1QPsi}),
assuming $\alpha\left(0\right)=\alpha\left(\infty\right)=0$, we obtain
\begin{equation}
\mathcal{L}^{-1}\left\{ Q\Psi\right\} \left(L\right)=Q\mathcal{L}^{-1}\left\{ \Psi\right\} \left(L\right)-e^{LK}\partial_{L}\left(e^{-LK}\alpha\left(L\right)\right)\,,\label{eq:tildeQPsi}
\end{equation}
where
\begin{equation}
\alpha\left(L\right)\equiv\mathcal{L}^{-1}\left\{ Fc\frac{K}{1-F^{2}}cF\right\} \left(L\right)\,.\label{eq:alpha}
\end{equation}

We expect $\alpha\left(\infty\right)=0$ for regular solutions. $\alpha\left(0\right)$
is related to the behavior of $F\left(K\right),\frac{K}{1-F^{2}}$
for $K\sim\infty$ and may not vanish even if $\Psi$ is regular.
For example, the Erler-Schnabl solution \cite{Erler:2009uj} has 
\begin{eqnarray*}
f\left(L\right) & = & \frac{1}{\Gamma\left(\frac{1}{2}\right)}L^{-\frac{1}{2}}e^{-L}\,,\\
\alpha\left(L\right) & = & e^{-L}\frac{1}{\left(\Gamma\left(\frac{1}{2}\right)\right)^{2}}\int_{0}^{L}dL^{\prime}\left(L-L^{\prime}\right)^{-\frac{1}{2}}L^{\prime-\frac{1}{2}}c\partial c\left(L^{\prime}\right)\,,
\end{eqnarray*}
 and 
\[
\alpha\left(0\right)=c\partial c\left(0\right)\,,
\]
With $\alpha\left(0\right)\ne0$, (\ref{eq:tildeQPsi}) cannot be
valid for such solutions.

In order to get an identity similar to (\ref{eq:tildeQPsi}) for the
solutions with $\alpha\left(\infty\right)=0,\alpha\left(0\right)\ne0$,
we regularize $\Psi$ and consider
\[
\Psi_{\eta}\equiv F\left(K\right)e^{-\eta K}c\frac{BK}{1-F^{2}\left(K\right)}e^{-\eta K}cF\left(K\right)e^{-\eta K}\,,
\]
for $\eta>0$. $\Psi_{\eta}$ coincides with the original one in the
limit $\eta\to0$ and 
\begin{eqnarray*}
\mathcal{L}^{-1}\left\{ \Psi_{\eta}\right\} \left(L\right) & = & \int dL_{1}dL_{2}dL_{3}\delta\left(L-L_{1}-L_{2}-L_{3}\right)\\
 &  & \hphantom{+\int dL_{1}d}\times c\left(L_{2}+L_{3}\right)Bc\left(L_{3}\right)\mathcal{L}^{-1}\left\{ F_{\eta}\right\} \left(L_{1}\right)\mathcal{L}^{-1}\left\{ \tilde{F}_{\eta}\right\} \left(L_{2}\right)\mathcal{L}^{-1}\left\{ F_{\eta}\right\} \left(L_{3}\right)\,,
\end{eqnarray*}
where
\begin{eqnarray*}
F_{\eta}\left(K\right) & \equiv & F\left(K\right)e^{-\eta K}\,,\\
\tilde{F}_{\eta}\left(K\right) & \equiv & \frac{K}{1-F^{2}\left(K\right)}e^{-\eta K}\,.
\end{eqnarray*}
$\mathcal{L}^{-1}\left\{ F_{\eta}\right\} \left(L\right),\mathcal{L}^{-1}\left\{ \tilde{F}_{\eta}\right\} \left(L\right)$
vanish for $L<\eta$ and we do not encounter any problem in deriving
\begin{equation}
\mathcal{L}^{-1}\left\{ Q\Psi_{\eta}\right\} \left(L\right)=Q\mathcal{L}^{-1}\left\{ \Psi_{\eta}\right\} \left(L\right)-e^{LK}\partial_{L}\left(e^{-LK}\alpha_{\eta}\left(L\right)\right)\,,\label{eq:tildeQPsieta}
\end{equation}
where
\[
\alpha_{\eta}\left(L\right)\equiv\mathcal{L}^{-1}\left\{ F_{\eta}c\tilde{F}_{\eta}cF_{\eta}\right\} \left(L\right)\,.
\]
$\alpha_{\eta}\left(L\right)\sim\alpha\left(L\right)$ for $L\gg\eta$
and $\alpha_{\eta}\left(L\right)=0$ for $L<3\eta$. Therefore, in
the limit $\eta\to0$, 
\[
\partial\alpha_{\eta}\left(L\right)\to\partial\alpha\left(L\right)+\delta\left(L\right)\alpha\left(0\right)\,,
\]
and (\ref{eq:tildeQPsieta}) becomes
\begin{equation}
\mathcal{L}^{-1}\left\{ Q\Psi\right\} \left(L\right)=Q\mathcal{L}^{-1}\left\{ \Psi\right\} \left(L\right)-e^{LK}\partial_{L}\left(e^{-LK}\alpha\left(L\right)\right)-\delta\left(L\right)\alpha\left(0\right)\,,\label{eq:tildeQPsi2}
\end{equation}
which can be used for solutions with $\alpha\left(\infty\right)=0,\alpha\left(0\right)\ne0$,
provided $\alpha\left(0\right)$ is well-defined. One can check that
the Laplace transform of the right hand side yields $Q\Psi$.

\section{Correlation functions of $X$ variables \label{sec:Correlation-functions-of}}

In the calculations in section \ref{sec:KBc}, we need the correlation
functions of $X$ variables, which are described by the free worldsheet
theory with the Neumann boundary condition, on $C_{L}$. A conformal
transformation which maps $C_{L}$ to the upper half plane is given
as 
\begin{eqnarray*}
C_{L} & \to & \mathrm{UHP}\\
z & \to & \xi=\tan\frac{\pi z}{L}\,.
\end{eqnarray*}
From the correlation functions 
\begin{eqnarray*}
\left\langle \partial X^{\mu}\left(\xi\right)\partial X^{\nu}\left(\xi^{\prime}\right)\right\rangle _{\mathrm{UHP}} & = & \frac{-\frac{1}{2}\eta^{\mu\nu}}{\left(\xi-\xi^{\prime}\right)^{2}}\,,\\
\left\langle \partial X^{\mu}\left(\xi\right)\bar{\partial}X^{\nu}\left(\bar{\xi}^{\prime}\right)\right\rangle _{\mathrm{UHP}} & = & \frac{-\frac{1}{2}\eta^{\mu\nu}}{\left(\xi-\bar{\xi}^{\prime}\right)^{2}}\,,
\end{eqnarray*}
we can get
\begin{eqnarray}
\left\langle \partial X^{\mu}\left(z\right)\partial X^{\nu}\left(z^{\prime}\right)\right\rangle _{C_{L}} & = & -\frac{1}{2}\eta^{\mu\nu}\left(\frac{\pi}{L}\right)^{2}\frac{1}{\sin^{2}\frac{\pi\left(z-z^{\prime}\right)}{L}}\,,\nonumber \\
\left\langle \partial X^{\mu}\left(z\right)\bar{\partial}X^{\nu}\left(\bar{z}^{\prime}\right)\right\rangle _{C_{L}} & = & -\frac{1}{2}\eta^{\mu\nu}\left(\frac{\pi}{L}\right)^{2}\frac{1}{\sin^{2}\frac{\pi\left(z-\bar{z}^{\prime}\right)}{L}}\,.\label{eq:XXCL}
\end{eqnarray}

We are interested in the correlation function of the form $\left\langle \left(X^{0}\left(z,\bar{z}\right)-X^{0}\left(z_{0},\bar{z}_{0}\right)\right)\partial X^{0}\left(z\right)\right\rangle _{C_{L}}$.
Since the difference $X^{0}\left(z,\bar{z}\right)-X^{0}\left(z_{0},\bar{z}_{0}\right)$
for some $z_{0},\bar{z}_{0}$ can be written as 
\[
X^{0}\left(z,\bar{z}\right)-X^{0}\left(z_{0},\bar{z}_{0}\right)=\int_{z_{0}}^{z}dz^{\prime}\partial X^{0}\left(z^{\prime}\right)+\int_{\bar{z}_{0}}^{\bar{z}}d\bar{z}^{\prime}\bar{\partial}X^{0}\left(\bar{z}^{\prime}\right)\,,
\]
using $\partial X^{0},\bar{\partial}X^{0}$, the correlation function
$\left\langle \left(X^{0}\left(z,\bar{z}\right)-X^{0}\left(z_{0},\bar{z}_{0}\right)\right)\partial X^{0}\left(z\right)\right\rangle _{C_{L}}$
is well-defined. Here it is assumed that the operators are normal
ordered as 
\begin{equation}
:X^{0}\partial X^{0}:\left(z,\bar{z}\right)\equiv\lim_{z^{\prime}\to z}\left[X^{0}\left(z,\bar{z}\right)\partial X^{0}\left(z^{\prime}\right)-\frac{1}{2}\frac{1}{z^{\prime}-z}\right]\,.\label{eq:normalorder}
\end{equation}
From (\ref{eq:XXCL}) we obtain 
\begin{eqnarray}
 &  & \left\langle \left(X^{0}\left(z,\bar{z}\right)-X^{0}\left(z_{0},\bar{z}_{0}\right)\right)\partial X^{0}\left(z\right)\right\rangle _{C_{L}}\nonumber \\
 &  & \qquad=\frac{\pi}{2L}\left[\cot\frac{\pi\left(z-\bar{z}\right)}{L}-\cot\frac{\pi\left(z-z_{0}\right)}{L}-\cot\frac{\pi\left(z-\bar{z}_{0}\right)}{L}\right]\,.\label{eq:X-X0CL}
\end{eqnarray}
If one chooses the reference point $z_{0}$ to be $i\infty$, we get
\[
\left\langle \left(X^{0}\left(z,\bar{z}\right)-X^{0}\left(i\infty,-i\infty\right)\right)\partial X^{0}\left(z\right)\right\rangle _{C_{L}}=\frac{\pi}{2L}\cot\frac{\pi\left(z-\bar{z}\right)}{L}\,.
\]

\section{Derivation of (\ref{eq:RN})(\ref{eq:GPsiGamma})\label{sec:Calculation-of-the}}

We would like to calculate the second and the third terms on the right
hand side of (\ref{eq:anomaly2}) in the limit $\epsilon\to0$. These
can be calculated basically using the $s$-$z$ trick \cite{Murata:2011ex,Murata:2011ep}. 

\begin{figure}
\begin{centering}
\includegraphics[scale=0.6]{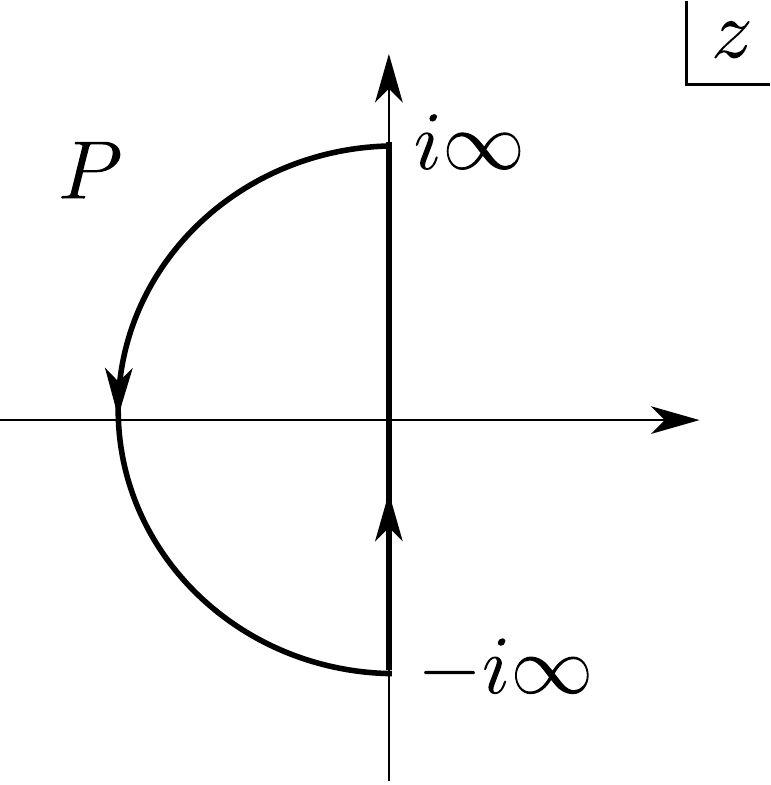}
\par\end{centering}

\caption{contour $P$\label{fig:contour}}
\end{figure}

Using 
\begin{eqnarray*}
\mathcal{L}^{-1}\left\{ \Gamma_{\epsilon}\right\} \left(L\right) & = & \int_{0}^{\infty}dL_{1}dL_{2}\delta\left(L-\sum_{i}L_{i}\right)c(L_{2})c(0)\mathcal{L}^{-1}\left\{ F_{\epsilon}^{2}\right\} (L_{1})\mathcal{L}^{-1}\left\{ \frac{K+\epsilon}{G_{\epsilon}}\right\} (L_{2})\,,
\end{eqnarray*}
and
\begin{eqnarray*}
 &  & \left\langle c\left(L_{2}\right)c\left(0\right)c\left(z\right)\right\rangle _{C_{L}}\\
 &  & \qquad=-\frac{1}{2}\left(\frac{L}{\pi}\right)^{3}\left[\left(\sin\left(\frac{\pi z}{L}\right)\right)^{2}\sin\frac{2\pi L_{2}}{L}-\left(\sin\left(\frac{\pi L_{2}}{L}\right)\right)^{2}\sin\frac{\pi z}{L}\right]\,,\\
 &  & \left\langle c\left(L_{2}\right)c\left(0\right)\left(\int_{i\delta}^{i\Lambda}\frac{dz}{2\pi i}4\partial X^{0}\left(z\right)\bar{c}\bar{\partial}X^{0}\left(\bar{z}\right)-\int_{-i\delta}^{-i\Lambda}\frac{d\bar{z}}{2\pi i}4\bar{\partial}X^{0}\left(\bar{z}\right)c\partial X^{0}\left(z\right)\right)\right\rangle _{C_{L}}\\
 &  & \qquad\underset{(\delta,\Lambda)\to(0,\infty)}{\longrightarrow}\frac{1}{4\pi}\left(\frac{L}{\pi}\right)^{2}\sin\frac{2\pi L_{2}}{L}\,,\\
 &  & \langle c\left(L_{2}\right)c\left(0\right)\kappa\left(i\delta,-i\delta\right)\rangle_{C_{L}}\underset{\delta\to0}{\longrightarrow}0\,,
\end{eqnarray*}
$\left\langle I\right|\chi\left|\Gamma_{\epsilon}\right\rangle $
becomes
\begin{eqnarray}
\left\langle I\right|\chi\left|\Gamma_{\epsilon}\right\rangle  & = & \frac{-1}{4\pi^{3}}\epsilon\int_{0}^{\infty}dss^{2}\int_{0}^{\infty}dL_{1}dL_{2}\delta\left(s-\sum_{i}L_{i}\right)\nonumber \\
 &  & \hphantom{\frac{-1}{4\pi^{3}}\epsilon\int_{0}^{\infty}dL}\times\mathcal{L}^{-1}\left\{ G_{\epsilon}\right\} (L_{1})\mathcal{L}^{-1}\left\{ \frac{K+\epsilon}{G_{\epsilon}}\right\} (L_{2})\sin\frac{2\pi}{s}L_{2}\nonumber \\
 & = & \frac{-1}{4\pi^{3}}\epsilon\int_{0}^{\infty}dss^{2}\int_{0}^{\infty}dL_{1}dL_{2}\int_{-i\infty}^{i\infty}\frac{dz}{2\pi i}e^{\left(s-\sum_{i}L_{i}\right)z}\nonumber \\
 &  & \hphantom{\frac{-1}{4\pi^{3}}\epsilon\int_{0}^{\infty}dL}\times\mathcal{L}^{-1}\left\{ G_{\epsilon}\right\} (L_{1})\mathcal{L}^{-1}\left\{ \frac{K+\epsilon}{G_{\epsilon}}\right\} (L_{2})\sin\frac{2\pi}{s}L_{2}\nonumber \\
 & = & \frac{i}{8\pi^{3}}\epsilon\int_{0}^{\infty}dss^{2}\int_{-i\infty}^{i\infty}\frac{dz}{2\pi i}e^{sz}G_{\epsilon}(z)\Delta\left(\frac{z+\epsilon}{G_{\epsilon}}\right)\nonumber \\
 & = & \frac{i}{8\pi^{3}}\epsilon\int_{0}^{\infty}dss^{2}\oint_{P}\frac{dz}{2\pi i}e^{sz}G_{\epsilon}(z)\Delta\left(\frac{z+\epsilon}{G_{\epsilon}}\right)\,.\label{eq:Gammachi}
\end{eqnarray}
Here $P$ is contour on the $z$ plane shown in figure \ref{fig:contour}
and $\Delta$ is defined as \cite{Murata:2011ex,Murata:2011ep}
\[
\Delta F(z)=F\left(z-\frac{2\pi i}{s}\right)-F\left(z+\frac{2\pi i}{s}\right).
\]
For the Murata-Schnabl solution (\ref{eq:MurataSchnabl}), (\ref{eq:Gammachi})
is evaluated as
\begin{eqnarray}
\left\langle I\right|\chi\left|\Gamma_{\epsilon}\right\rangle  & = & R_{N}+\mathcal{O}\left(\epsilon\right),\label{eq:epsilon1-1}\\
R_{N} & \equiv & \begin{cases}
-\frac{i}{8\pi^{3}}\sum_{k=0}^{N-2}\frac{N!}{k!\left(k+2\right)!\left(N-2-k\right)!}\left((2\pi i)^{k+2}-\left(-2\pi i\right)^{k+2}\right) & ,\ \left(N\geq1\right)\,,\\
\frac{i}{8\pi^{3}}\sum_{k=0}^{-N-1}\frac{\left(1-N\right)!}{k!\left(k+2\right)!\left(-N-1-k\right)!}\left((2\pi i)^{k+2}-\left(-2\pi i\right)^{k+2}\right) & ,\ \left(N\leq0\right)\,,
\end{cases}\nonumber 
\end{eqnarray}
for $\epsilon\ll1$. 

The third term on the right hand side of (\ref{eq:anomaly2}) becomes
\begin{eqnarray*}
 &  & \int dL_{1}dL_{2}\frac{L_{1}}{L_{1}+L_{2}}\langle e^{L_{2}K}\mathcal{L}^{-1}\left\{ \Psi_{\epsilon}\right\} \left(L_{1}\right)e^{-L_{2}K}\mathcal{L}^{-1}\left\{ \Gamma_{\epsilon}\right\} \left(L_{2}\right)\rangle_{C_{L_{1}+L_{2}}}\\
 &  & \qquad=\epsilon\int_{0}^{\infty}ds\prod_{i=1}^{4}dL_{i}\delta\left(s-\sum_{i=1}^{4}L_{i}\right)\frac{L_{1}+L_{2}}{s}\\
 &  & \hphantom{\qquad=\epsilon}\times\mathrm{Tr}\left[e^{-L_{1}K}\mathcal{L}^{-1}\left\{ G_{\epsilon}\right\} \left(L_{1}\right)cBe^{-L_{2}K}\mathcal{L}^{-1}\left\{ \frac{K+\epsilon}{G_{\epsilon}}\right\} \left(L_{2}\right)c\right.\\
 &  & \hphantom{\qquad=\epsilon e^{-L_{1}K}\mathcal{L}^{-1}\left\{ G_{\epsilon}\right\} }\times\left.e^{-L_{3}K}\mathcal{L}^{-1}\left\{ G_{\epsilon}\right\} \left(L_{3}\right)ce^{-L_{4}K}\mathcal{L}^{-1}\left\{ \frac{K+\epsilon}{G_{\epsilon}}\right\} \left(L_{4}\right)c\right]\,.
\end{eqnarray*}
 Using
\[
L\mathcal{L}^{-1}\left\{ f\right\} \left(L\right)=\mathcal{L}^{-1}\left\{ \partial f\right\} \left(L\right)\,,
\]
and eq.(2.5) in \cite{Murata:2011ep}, we obtain
\begin{eqnarray*}
 &  & \int dL_{1}dL_{2}\frac{L_{1}}{L_{1}+L_{2}}\langle e^{L_{2}K}\mathcal{L}^{-1}\left\{ \Psi_{\epsilon}\right\} \left(L_{1}\right)e^{-L_{2}K}\mathcal{L}^{-1}\left\{ \Gamma_{\epsilon}\right\} \left(L_{2}\right)\rangle_{C_{L_{1}+L_{2}}}\\
 &  & \qquad=\frac{i}{8\pi^{3}}\epsilon\int_{0}^{\infty}dss\oint_{C}\frac{dz}{2\pi i}e^{sz}\frac{1}{2i}\\
 &  & \hphantom{\qquad=\quad}\times\left\{ \left[\frac{z+\epsilon}{G_{\epsilon}},G_{\epsilon},\frac{z+\epsilon}{G_{\epsilon}},G_{\epsilon}^{\prime}\right]+\left[\left(\frac{z+\epsilon}{G_{\epsilon}}\right)^{\prime},G_{\epsilon},\frac{z+\epsilon}{G_{\epsilon}},G_{\epsilon}\right]\right\} ,
\end{eqnarray*}
where
\begin{eqnarray*}
\left[F_{1},F_{2},F_{3},F_{4}\right] & \equiv & \left[-F_{1}\Delta F_{2}F_{3}F_{4}^{\prime}+F_{1}\Delta\left(F_{2}F_{3}^{\prime}\right)F_{4}+F_{1}\Delta\left(F_{2}F_{3}\right)F_{4}^{\prime}-F_{1}F_{2}^{\prime}F_{3}\Delta F_{4}\right.\\
 &  & \left.+F_{1}F_{2}^{\prime}\Delta\left(F_{3}F_{4}\right)+F_{1}F_{2}\Delta\left(F_{3}^{\prime}F_{4}\right)-F_{1}\Delta\left(F_{2}F_{3}^{\prime}F_{4}\right)-F_{1}\left(F_{2}\Delta F_{3}F_{4}\right)^{\prime}\right].
\end{eqnarray*}
The contribution of $\mathcal{O}\left(\epsilon^{0}\right)$ is given
by the following replacements
\begin{eqnarray*}
G^{\prime}\left(z\right) & \to & -\left(N-1\right)G\left(z\right)\,,\\
G^{\prime\prime}\left(z\right) & \to & N\left(N-1\right)\frac{1}{z^{2}}G\left(z\right)\,,\\
\left(\frac{z}{G}\right)^{\prime}\left(z\right) & \to & NG^{-1}\left(z\right)\,,
\end{eqnarray*}
and one can see 
\[
\int dL_{1}dL_{2}\frac{L_{1}}{L_{1}+L_{2}}\langle e^{L_{2}K}\mathcal{L}^{-1}\left\{ \Psi_{\epsilon}\right\} \left(L_{1}\right)e^{-L_{2}K}\mathcal{L}^{-1}\left\{ \Gamma_{\epsilon}\right\} \left(L_{2}\right)\rangle_{C_{L_{1}+L_{2}}}\sim\mathcal{O}\left(\epsilon\right)\,.
\]

\bibliographystyle{utphys}
\bibliography{private,book8-5}

\end{document}